\documentclass[aps,twocolumn,showpacs,preprintnumbers,nofootinbib,prd,10pt,superscriptaddress]{revtex4-1}

\usepackage{graphicx} 
\usepackage{amssymb}
\usepackage{mathrsfs}
\usepackage{bm}
\usepackage{amsmath}
\usepackage{multirow}
\usepackage[usenames,dvipsnames]{xcolor}
\usepackage{hyperref}
\usepackage{verbatim}
\usepackage{orcidlink}
\usepackage{soul}
\hypersetup{colorlinks=true, citecolor=Purple, linkcolor=Purple,
urlcolor=Purple}

\newcommand{\dd}{\mathrm{d}}
\newcommand{\derivative}[2]{\frac{\dd #1}{\dd #2}}
\newcommand{\lmo}[1]{#1_{\ell m \omega}}
\newcommand{\derr}[2]{\frac{\mathrm{d}^2 #1}{\mathrm{d} #2 ^2}}

\definecolor{darkgreen}{rgb}{0,0.5,0}

\def\beq{\begin{eqnarray}}
\def\eeq{\end{eqnarray}}
\def\nn{\nonumber}
\def\be{\begin{equation}}
\def\ee{\end{equation}}

\renewcommand{\a}{\alpha}
\renewcommand{\b}{\beta}

\newcommand{\n}{\nu}
\renewcommand{\t}{\tau}
\newcommand{\z}{\omega}
\newcommand{\G}{\Gamma}

\newcommand\EN{\mathcal{E}}
\newcommand\ANG{\mathcal{L}}
\renewcommand{\Re}{{\rm Re }}
\renewcommand{\Im}{{\rm Im}}
\begin{document}
\title{Quasinormal ringing of Kerr black holes. III. Excitation coefficients for equatorial inspirals from the innermost stable circular orbit}

\author{Matteo Della Rocca \orcidlink{0009-0001-4470-3694}}
\email{matteo.dellarocca@phd.unipi.it}
\affiliation{Dipartimento di Fisica, Sapienza Universit\`a di Roma, Piazzale Aldo Moro 5, 00185, Roma, Italy}
\affiliation{Dipartimento di Fisica, Universit\`a di Pisa, Largo B. Pontecorvo 3, 56127 Pisa, Italy}
\affiliation{INFN, Sezione di Pisa, Largo B. Pontecorvo 3, 56127 Pisa, Italy}

\author{Laura Pezzella \orcidlink{0009-0008-6357-6417}}
\email{laura.pezzella@gssi.it}
\affiliation{Gran Sasso Science Institute (GSSI), I-67100 L’Aquila, Italy}
\affiliation{INFN, Laboratori Nazionali del Gran Sasso, I-67100 Assergi, Italy}
\affiliation{William H. Miller III Department of Physics and Astronomy, Johns Hopkins University, 3400 N. Charles Street, Baltimore, Maryland, 21218, USA}

\author{Emanuele Berti \orcidlink{0000-0003-0751-5130}}
 \email{berti@jhu.edu}
\affiliation{William H. Miller III Department of Physics and Astronomy, Johns Hopkins University, 3400 N. Charles Street, Baltimore, Maryland, 21218, USA}
\author{Leonardo Gualtieri \orcidlink{0000-0002-1097-3266}}
 \email{leonardo.gualtieri@unipi.it}
\affiliation{Dipartimento di Fisica, Universit\`a di Pisa, Largo B. Pontecorvo 3, 56127 Pisa, Italy}
\affiliation{INFN, Sezione di Pisa, Largo B. Pontecorvo 3, 56127 Pisa, Italy}
\author{Andrea Maselli \orcidlink{0000-0001-8515-8525}}
\email{andrea.maselli@gssi.it}
\affiliation{Gran Sasso Science Institute (GSSI), I-67100 L’Aquila, Italy}
\affiliation{INFN, Laboratori Nazionali del Gran Sasso, I-67100 Assergi, Italy}
\begin{abstract} 
The remnant of a black hole binary merger settles into a stationary configuration by ``ringing down'' through the emission of gravitational waves that consist of a superposition of damped exponentials with discrete complex frequencies -- the remnant black hole's quasinormal modes. While the frequencies themselves depend solely on the mass and spin of the remnant, the mode amplitudes depend on the merger dynamics. We investigate quasinormal mode excitation by a point particle plunging from the innermost stable circular orbit of a Kerr black hole. Our formalism is general, but we focus on computing the quasinormal mode excitation coefficients in the frequency domain for equatorial orbits, and we analyze their dependence on 
the remnant black hole spin. We find that higher overtones and subdominant multipoles of the radiation become increasingly significant for rapidly rotating black holes. This suggests that the prospects for detecting overtones and higher-order modes are considerably enhanced for highly spinning merger remnants.
\end{abstract}

\maketitle
\section{Introduction}

%
The modeling of the signal from merging black hole (BH) binaries is a delicate task, requiring an interplay of numerical simulations and perturbative solutions of the Einstein equations. 
Hybrid models combining analytical and numerical methods are needed both because numerical simulations are computationally costly, and because it is difficult to cover the whole parameter space, including binaries with large mass ratios, precessing spins, and nonzero eccentricity (see e.g.~\cite{Ajith:2007kx,Nagar:2018zoe,Estelles:2021gvs,Ramos-Buades:2023ehm}).

The ringdown signal following the merger is in principle simpler, as it can be described with good accuracy as a superposition of complex exponentials corresponding to the quasinormal modes (QNMs) of the final BH~\cite{Kokkotas:1999bd,Berti:2009kk,Berti:2025hly}:
\begin{equation}
    h(t,r,\theta,\phi)\simeq \frac{1}{r}\sum_{\ell mn} C_{\ell m n}\,_{-2}S_{\ell m}^{a\omega_{\ell mn}}(\theta,\phi)e^{-i\omega_{\ell m n}(t-r_\star)}\,.
    \label{eq:mode_expansion}
\end{equation}
Here $h=h_+-i\,h_\times$ is the gravitational strain,
$\omega_{\ell m n}$ denotes the complex QNM frequencies for a Kerr BH with mass $M$ and spin parameter $a$, $r$ is the (luminosity) distance to the source, $r_\star$ is the tortoise coordinate, and the quantities $C_{\ell m n}$ are known as the QNM ``excitation coefficients''.
The functions $\,_{-2}S_{\ell m}^{a\omega}$ are the spin-weighted spheroidal harmonics with indices $\ell,m$ used to separate the angular dependence~\cite{Teukolsky_1973} and $n$ is the overtone index, sorting the modes by the magnitude of their imaginary part. In the following, we will often use the collective index $q=(\ell,m,n)$ to label a specific QNM. 
We remark that the QNMs are not a complete basis~\cite{Nollert:1998ys}, and at late times the mode expansion~\eqref{eq:mode_expansion} is dominated by a power-law tail contribution~\cite{Price:1971fb}.

A better theoretical understanding of ringdown excitation is necessary to accurately model the gravitational signal from BH binaries.
In particular, it is important to predict how the excitation coefficients $C_q\equiv C_{\ell m n}$---i.e., the (complex) amplitudes of the different modes in the QNM expansion~\eqref{eq:mode_expansion}---depend on the properties of the merger progenitors~\cite{Berti:2005ys,Berti:2006hb,Buonanno:2006ui,Berti:2007zu,Berti:2007fi,Kamaretsos:2011um,Kamaretsos:2012bs}. The main goal of this paper is to compute these quantities for extreme mass-ratio binaries using BH perturbation theory. This information is complementary to the results of numerical relativity simulations, and it can be used to calibrate and validate phenomenological ringdown models, such as those based on fits of comparable-mass numerical relativity simulations~\cite{Baibhav:2017jhs,Giesler:2019uxc,Li:2021wgz,Ma:2021znq,Ma:2022wpv,Baibhav:2023clw,Cheung:2023vki,Takahashi:2023tkb,Zhu:2023fnf,MaganaZertuche:2024ajz,Giesler:2024hcr,Mitman:2025hgy,Morisaki:2025gyu,Crescimbeni:2025ytx}. 
In addition, recent work suggests that perturbative results are remarkably accurate even for intermediate or comparable mass-ratio binaries~\cite{vandeMeent:2020xgc,Wardell:2021fyy,Albertini:2022rfe}.

Even the simple problem of computing how QNMs are excited by plunging particles is not fully understood.  
This line of research started four decades ago with seminal work by Leaver~\cite{Leaver:1986gd}, who computed the excitation coefficients for a particle falling radially into a Schwarzschild BH. Leaver's results were later extended to particles plunging radially along the symmetry axis of a Kerr BH~\cite{Berti:2006wq,Zhang:2013ksa}, and to the excitation of modes by localized initial data~\cite{Kubota:2025hjk}. 
The ringdown amplitudes and excitation coefficients for particles plunging from the innermost stable circular orbit (ISCO) are more observationally relevant than radial plunges. These quantities have been computed in the case of Schwarzschild BHs by various authors~\cite{Hadar:2009ip,Hadar:2011vj,Folacci:2018cic,OuldElHadj:2024psw}.
Following up on early work relating QNMs to the properties of null geodesics in the Kerr spacetime~\cite{1972ApJ...172L..95G,Cardoso:2008bp,Dolan:2010wr,Yang:2012he}, Hughes and collaborators studied
how the relative QNM amplitudes in the extreme mass-ratio limit encode information about the binary's geometry---including the misalignment between the BH spin and the small body's orbit, the eccentricity, and the orbital anomaly---using time-domain evolution codes~\cite{Hughes:2019zmt,Apte:2019txp,Lim:2019xrb,Lim:2022veo,Becker:2024xdi,Becker:2025zzw}.
Oshita and collaborators extended Leaver's work to higher overtones~\cite{Oshita:2021iyn,Oshita:2024wgt} and near-extremal Kerr BHs~\cite{Oshita:2022yry,Watarai:2024vni}, highlighting interesting connections between QNM excitation and the hole's greybody factors~\cite{Oshita:2022pkc,Oshita:2023cjz,Okabayashi:2024qbz}, and the possibility that ``direct waves''  modulated by the plunging motion may reveal observational imprints of the remnant's horizon properties~\cite{Lu:2025vol,Oshita:2025qmn}. 

In realistic physical scenarios, the modeling of the full coalescence requires a good understanding of the transition between the inspiral phase (traditionally assumed to terminate at the ISCO) and the plunge phase. Different models have been proposed, and they provide different prescriptions for how the orbital parameters change in the inspiral-to-plunge transition for finite values of the mass ratio~\cite{Buonanno:2000ef,Ori:2000zn,Kesden:2011ma,Compere:2021iwh,Lhost:2024jmw}.
In Ref.~\cite{Watarai:2024huy}, the excitation coefficients in the Kerr background were extracted from a numerical time evolution in which the transition from inspiral to plunge was described by the Ori-Thorne model~\cite{Ori:2000zn}.  In Ref.~\cite{DeAmicis:2025xuh}, similar models~\cite{Buonanno:2000ef,Ori:2000zn} were used to compute the mode amplitudes during the plunge using a numerical evolution.
Other recent efforts to model ringdown excitation in the extreme mass-ratio limit include Ref.~\cite{Kuchler:2025hwx}, that applied gravitational self-force models based on a slow ``offline'' stage (in which waveform ingredients are precomputed as functions of the orbital phase space) and a fast ``online'' stage (where the waveform is generated by evolving through the phase space) to
quasicircular, nonspinning binaries; and Ref.~\cite{Honet:2025dho}, where the same next-to-next-to-leading-order transition-to-plunge waveform model was generalized to include the spin of the primary BH.

Models for the transition from inspiral to plunge are difficult to extend to generic orbits without introducing arbitrary parameters~\cite{Apte:2019txp}. There are indications~\cite{Lhost:2024jmw,Faggioli:2025hff} that at least in the small mass-ratio limit this transition is well described by the so-called {\it critical plunge geodesics}~\cite{Mummery:2022ana,Dyson:2023fws}: a class of geodesics starting from the last stable orbit (or from the ISCO in the circular, equatorial case).
In this work we will compute QNM excitation using the critical plunge geodesics found in Ref.~\cite{Mummery:2022ana}, where---building on early results by Chandrasekhar~\cite{Chandrasekhar:1985kt}---it was noted that the constants of motion for the circular, equatorial ISCO of the Kerr spacetime also describe  a plunging geodesic starting from the ISCO at asymptotic past, and that these geodesics have simple analytical expressions. These results were extended to generic timelike plunge geodesics in Ref.~\cite{Dyson:2023fws}: in this case, the critical geodesic can be expressed in terms of elliptic integrals.  Because critical plunge geodesics are known for generic timelike orbits, accurate in the extreme mass-ratio limit, and known in closed form, they are ideal for a systematic investigation of QNM excitation.
We will work in the frequency domain, extending the formalism developed in~\cite{Zhang:2013ksa}.
To validate our results, we will compare ``pure ringdown'' gravitational waveforms of the form of Eq.~\eqref{eq:mode_expansion}, computed from the excitation coefficients, against the ``full'' waveforms computed using Green's function techniques (see e.g.~\cite{Nakamura:1987zz,Berti:2010ce}, and~\cite{Silva:2023cer,Yin:2025kls} for more recent work).

It is important to point out that there are several equivalent definitions of the excitation coefficients in the literature, because different formalisms are used to describe spacetime perturbations and different variables can be expanded in a form similar to Eq.~\eqref{eq:mode_expansion}~\cite{Leaver:1986gd,Berti:2006wq,Lo:2025njp,Kubota:2025hjk}. These conventions affect the comparison of the excitation coefficients for different modes.  Here we will focus on the gravitational strain, as this is the observable quantity, but note that the observed waveform also depends on the inclination of the source with respect to the observer through the angles $(\theta,\,\phi)$ in Eq.~\eqref{eq:mode_expansion}.

The paper is organized as follows. In Section~\ref{sec:theorysetup}
we introduce the theoretical framework and define the quantities needed to compute the excitation coefficients of QNMs in the presence of a generic point particle source.
In Section~\ref{sec:sourceterm} we derive the source term for a particle on a critical plunge geodesic.
In Section~\ref{sec:results} we present our main results, showing the excitation coefficients as functions of the BH dimensionless spin $a/M$.
To improve readability, some technical details are relegated to the Appendices.
Throughout this paper, unless stated otherwise, we use geometrical units ($G=c=1$). 

\section{Theoretical setup}\label{sec:theorysetup}

In this section we review the definition of the excitation coefficients and their calculation in the Kerr background.
We consider the spacetime describing a Kerr BH of mass $M$ and angular momentum $Ma$ in Boyer-Lindquist coordinates $(t,r,\theta,\phi)$~\cite{Boyer:1966qh}. In these coordinates, the location of the outer and inner horizons is given by the roots of the function $\Delta=r^2+a^2-2Mr$, i.e., $r_\pm=M\pm \sqrt{M^2-a^2}$. 

We start from the Teukolsky equation in the frequency domain (Sec.~\ref{sec:Teuk}), and then we introduce the Sasaki-Nakamura (SN) formalism (Sec.~\ref{sec:Sasaki_Nakamura_equation}). We isolate the QNM contribution by analytical continuation of a particular solution of the SN equation to complex frequencies (Sec.~\ref{sec:excitationcoefficients}), and finally we define the excitation coefficients for a generic source (Sec.~\ref{sec:hexcitationcoefficients}). 

\subsection{The Teukolsky equation}\label{sec:Teuk}

We study perturbations of Kerr BHs in the Newman-Penrose formalism~\cite{Newman:1961qr}. In this framework, the perturbations are encoded in a single quantity, the Weyl scalar $\psi_4$, which is related to the gravitational strain by
\begin{equation}
    \psi_4(t,r,\theta,\phi)=\frac{1}{2}\ddot h(t,r,\theta,\phi)
    \label{eq:psi4h}
\end{equation}
and satisfies the Teukolsky equation
\begin{equation}
\mathcal{J}(\psi_4)=4\pi T\ ,
\end{equation}
where $\mathcal{J}$ is a second-order differential operator acting on $\psi_4$, and $T$ is a source term~\cite{Teukolsky_1973}. This equation is separable in terms of an orthonormal set of angular functions, the spin-weighted spheroidal harmonics $\,_{-2}S^{a\omega}_{\ell m}$ (here defined following the conventions of~\cite{Breuer:1975book,Mino:1997bx}), where $\ell=2,3,\ldots$ and $m=-\ell,\ldots,\ell$. By expanding $\psi_4$ and the source as
\be
\rho^{-4}\psi_4=\frac{1}{\sqrt{2\pi}}\sum_{\ell m}\int\dd\omega \  e^{-i\omega t} \,_{-2}S^{a\omega}_{\ell m} R_{\ell m\omega}
\label{eq:expansion_psi}
\ee
and
\begin{equation}
    T=\frac{1}{\sqrt{2\pi}}\sum_{\ell m}\int \dd \omega \ e^{-i\omega t} \,_{-2}S^{a\omega}_{\ell m}T_{\ell m\omega}\; ,
\end{equation}
where $\rho=(r-i a \cos\theta)^{-1}$, the radial and angular parts of the perturbation decouple, and satisfy two distinct equations:
\begin{equation}
    \left[\Delta^{2}\frac{\dd }{\dd r}\left(\frac{1}{\Delta}\frac{\dd \ }{\dd r}\right)-V\right] R_{\ell m \omega}=T_{\ell m\omega}\, \label{eq:R_teukolsky}\ ,
\end{equation}
and 
\begin{align}
\label{eq:S_teukolsky}
\frac{\dd}{\dd z}\bigg[(1-z^2)\frac{\dd}{\dd z} \,_{-2}S^{a\omega}_{\ell m}\bigg]&+\bigg[a^2\omega^2z^2+4a\omega z+\mathcal{A}_{\ell m}\\ \nonumber
&-\frac{(m-2z)^2}{1-z^2}-2\bigg]\,_{-2}S^{a\omega}_{\ell m}=0\, ,
\end{align} 
where $z=\cos \theta$, 
\begin{equation}
    V=-\frac{K^2+4i(r-M)K}{\Delta}+8i\omega r+\lambda \ ,
    \label{eq:TeukolskyPotential}
\end{equation}
$K=(r^2+a^2)\omega-a m$, and $\lambda=\mathcal{A}_{\ell m }+a^2\omega^2-2am\omega$.

\subsection{Sasaki-Nakamura equation}\label{sec:Sasaki_Nakamura_equation}

The scattering potential $V(r)$ appearing in the Teukolsky equation \eqref{eq:R_teukolsky} has a long-range behavior, which complicates numerical integration. To address this, we use the SN formalism~\cite{Sasaki:1981a,Sasaki:1982,Sasaki:1981sx,Nakamura:1987zz}, which transforms the Teukolsky equation into an equivalent differential equation (the SN equation) with a short-range potential~\cite{Sasaki:1981sx}. 
In the limit $a\to 0$, the SN equation reduces to the Regge-Wheeler and Zerilli equations for Schwarzschild perturbations (the Zerilli equation can be found through the transformation discussed in~\cite{Chandrasekhar:1985kt}). 

To derive the SN equation, we first consider the most general transformation of the function $R(r)$ that preserves the linear structure of Eq.~\eqref{eq:R_teukolsky}:
\begin{equation}
\chi_{\ell m\omega}(r)=\alpha(r) R_{\ell m\omega}(r)+\frac{\beta(r)}{\Delta}R'_{\ell m\omega}(r)\; ,\label{eq:chi_def}
\end{equation}
where $\alpha$ and $\beta$ are generic functions of $r$. We denote with a prime the derivative with respect to the radial coordinate $r$. 
Differentiating $\chi_{\ell m\omega}$ with respect to $r$ and using the Teukolsky equation, we obtain
\begin{equation}
    \Delta^2\left(\frac{1}{\Delta}\chi_{\ell m\omega}'\right)'-\Delta F\chi'_{\ell m\omega}-U\chi_{\ell m\omega}=\mathscr{S}_{\ell m\omega} \; ,\label{eq:chi_eq1}
\end{equation}
where the source term $\mathscr{S}_{\ell m\omega}$ will be specified later. Here, $F$ and $U$, which depend only on the radial coordinate, are given by
\begin{align}
F=&  \frac{\gamma'}{\gamma}\ ,\label{eq:F_SN}\\
U=& V+\frac{\Delta^2}{\beta}\left[\left(2\alpha+\frac{\beta'}{\Delta}\right)'-\frac{\gamma'}{\gamma}\left(\alpha+\frac{\beta'}{\Delta}\right)\right]\ , \label{eq:U_SN}
\end{align}
where
\begin{equation}
        \gamma=\alpha\left(\alpha+\frac{\beta'}{\Delta}\right)-\frac{\beta}{\Delta}\left(\alpha'+\frac{\beta}{\Delta^2}V\right)\label{eq:gamma_SN} \ .
\end{equation}
We then introduce the function
\begin{equation}
X_{\ell m\omega}=\frac{\sqrt{r^2+a^2}}{\Delta}\chi_{\ell m\omega} \ ,
\label{eq:X_def}
\end{equation}
and, using Eq.~\eqref{eq:chi_eq1}, we derive the  equation for $X_{\ell m \omega}$ in terms of the tortoise coordinate $r_\star$: 
\begin{equation}
    \frac{\dd^2 X_{\ell m\omega}}{\dd r_\star^2}-\mathcal{F}\frac{\dd X_{\ell m\omega}}{\dd r_\star}-\mathcal{U} X_{\ell m\omega}=\mathcal{S}_{\ell m\omega}
    \ ,\label{eq:X_eq1}
\end{equation}
where
\begin{equation}
    r_\star=r+\frac{2Mr_+}{r_+-r_-}\log\left(\frac{r-r_+}{2M}\right)-\frac{2Mr_-}{r_+-r_-}\log\left(\frac{r-r_-}{2M}\right) 
\end{equation}
and 
\begin{align}
    \mathcal{F}=&\frac{\Delta F}{r^2+a^2}=\frac{\gamma_{,r_\star}}{\gamma} \ ,\\
    \mathcal{U}=&\frac{\Delta U}{(r^2+a^2)^2}+G^2+\frac{\Delta G'}{r^2+a^2}-\frac{\Delta G F}{r^2+a^2} \ ,\\
    G=&-\frac{\Delta'}{r^2+a^2}+\frac{r\Delta}{(r^2+a^2)^2}\ ,\\
    \mathcal{S}_{\ell m\omega}=&\frac{\mathscr{S}_{\ell m\omega}}{(r^2+a^2)^{3/2}}\ .
\end{align}
The functions $\alpha$, $\beta$ and $\gamma$ can be chosen such 
that Eq.~\eqref{eq:X_eq1} reduces to the Regge-Wheeler equation in 
the limit $a=0$.  A common choice is
\begin{align}
    \alpha&=-i \frac{K}{\Delta^2}\beta+3 i K'+\lambda+\Delta \frac{6}{r^2}\ ,\\
    \beta&= \Delta \left(-2 i K +\Delta'-\frac{4\Delta}{r}\right)\ ,\\
    \gamma&=\sum_{i=0}^4 c_i r^{-i} \ ,\label{eq:gamma_ci_def}
\end{align}
where
\begin{align}
    c_0=&-12 a \omega  (a \omega -m)+\lambda  (\lambda +2)-12 i M \omega\,,  \label{eq:c_0}\\
    c_1=&8 i a (3 a \omega -\lambda  (a \omega -m))\,,\\
    c_2=&12 a^2 \left[1-2 (a \omega -m)^2\right]-24 i a M (a \omega -m)\,,\\
    c_3=&-24 a^2 M+24 i a^3 (a \omega -m)\,,\\
    c_4=&12 a^4\,.
\end{align}
With this choice of $\alpha$, $\beta$ and $\gamma$, the transformation~\eqref{eq:chi_def} is called the SN transformation, the function $X_{\ell m\omega}$ defined in Eq.~\eqref{eq:X_def} is the SN function, and Eq.~\eqref{eq:X_eq1} is the SN equation. Finally, by introducing the rescaled function 
\begin{equation}
    \tilde X_{\ell m\omega}=\frac{X_{\ell m\omega}}{\sqrt{\gamma}} \label{eq:tilde_chi_def}
\end{equation}
we get rid of the term proportional to the first derivative in Eq.~\eqref{eq:X_eq1}:
\begin{equation}
    \frac{\dd^2 \tilde X_{\ell m\omega}}{\dd r_\star^2}+\tilde {\mathcal{F}}\tilde X_{\ell m\omega}=\frac{\mathcal{S}_{\ell m\omega}}{\sqrt{\gamma}} \, ,\label{eq:SN_final}
\end{equation}
where $\tilde {\mathcal{F}}= \frac{\mathcal{F}_{,r_\star}}{2}-\frac{\mathcal{F}^2}{4}-\mathcal{U}$. 
So far we have not specified the source term $\mathcal{S}_{\ell m \omega}$, which can be determined in terms of $T_{\ell m \omega}$ by inverting  the SN transformation and seeking a solution of the inhomogeneous problem of the form
\begin{equation}
    R_{\ell m\omega}=\frac{1}{\gamma}\left[\left(\alpha+\frac{\beta'}{\Delta}\right)\chi_{\ell m\omega}-\frac{\beta}{\Delta}\chi'_{\ell m\omega}\right]+ \mathcal{Q}\mathcal{S}_{\ell m\omega}\ ,\label{eq:from_chi_to_R}
\end{equation}
where $\mathcal{Q}$ is  a function of $r$. Substituting Eq.~\eqref{eq:from_chi_to_R} into Eq.~\eqref{eq:R_teukolsky}, and using Eq.~\eqref{eq:chi_eq1},
we obtain
\begin{align}
    T_{\ell m \omega}=\Delta^2\left[\frac{1}{\Delta}\left(\mathcal{Q}\mathscr{S}'_{\ell m\omega}\right)\right]'&+\Delta^2\left(\frac{\beta\mathscr{S}_{\ell m\omega}}{\gamma \Delta^3}\right)'\nonumber\\
    &+\left(\frac{\alpha}{\gamma}-V \mathcal{Q}\right)\mathscr{S}_{\ell m\omega} \ .
\label{eq:T_S_link}
\end{align}
Following Ref.~\cite{Nakamura:1987zz}, we can choose $\mathcal{Q}=(r^2+a^2)^{3/2}/\gamma$ and introduce the auxiliary function
\begin{equation}
    W_{\ell m \omega}(r)=\mathcal{S}_{\ell m\omega}\frac{r^2(r^2+a^2)^{3/2}}{\gamma \Delta }e^{i\xi} \ .
\label{eq:W_def}
\end{equation}
Combining Eqs.~\eqref{eq:T_S_link} and \eqref{eq:W_def} yields
\begin{equation}
    W_{\ell m \omega}^{\prime\prime}=-\frac{r^2 T_{\ell m\omega}}{\Delta^2}
    e^{i\xi} \ ,
\label{eq:W_T_link}
\end{equation}
with
\begin{align}
    \xi(r)=\int \dd r \ \frac{K}{\Delta}=&\,\omega r+\frac{2M\omega r_+-a m}{r_+-r_-}\log\left(\frac{r-r_+}{2M}\right)\nonumber\\
    &-\frac{2M\omega r_--a m}{r_+-r_-}\log\left(\frac{r-r_-}{2M}\right)\; .
\end{align}
Given the source components $T_{\ell m \omega}$, we can solve Eq.~\eqref{eq:W_T_link} to find $W_{\ell m\omega}$ and then, by Eq.~\eqref{eq:W_def}, $\mathcal{S}_{\ell m\omega}$.
%

\subsection{The Sasaki-Nakamura excitation coefficients}\label{sec:excitationcoefficients}
A particular solution of the SN equation~\eqref{eq:SN_final} can be found through the Green's function method:
\begin{equation}
\tilde X_{\ell m \omega}^{p}(r_\star)=\int_\mathbb{R}\dd r_\star' \ \frac{\tilde X_{\ell m \omega}^{\infty}({r_\star}_{>})\tilde X_{\ell m \omega}^{r_+}({r_\star}_{<})}{\mathcal{W}_{\ell m \omega}}\frac{\mathcal{S}_{\ell m\omega}(r_\star')}{\sqrt{\gamma}}  \ ,
\label{eq:SN_complete_solution}
\end{equation}
where ${r_{\star}}_>=\max\{r_\star,r_\star'\}$, ${r_\star}_<=\min\{r_\star,r_\star'\}$, and the functions $\tilde X_{\ell m \omega}^{r_+}$ and $\tilde X_{\ell m \omega}^{\infty}$ are solutions of 
the homogeneous equation with purely ingoing boundary conditions at the horizon and purely outgoing boundary conditions at infinity, respectively, such that
\begin{equation}
    \tilde X_{\ell m \omega}^{r_+}\to \begin{cases} e^{-i p_\omega r_\star}& r\sim r_+\\
   \tilde A_{\ell m \omega}^{\rm in} e^{-i\omega r_\star}+\tilde A_{\ell m \omega}^{\rm out} e^{i\omega r_\star} & r\sim \infty
    \end{cases}
      \label{eq:rescaled_SN_rh}
\end{equation}
\begin{equation}
    \tilde X_{\ell m \omega}^{\infty}\to \begin{cases}  \tilde B_{\ell m \omega}^{\rm in} e^{-i p_\omega r_\star}+\tilde B_{\ell m \omega}^{\rm out} e^{i p_\omega r_\star}& r\sim r_+\\
 e^{i\omega r_\star} & r\sim \infty
    \end{cases}
    \label{eq:rescaled_SN_inf}
\end{equation}
with $p_\omega\equiv \omega -m \Omega$, $\Omega=a/\left(2Mr_+\right)$, and where
\begin{equation}
    \mathcal{W}_{\ell m \omega}=\tilde X_{\ell m \omega}^{r_+}\frac{\dd\tilde X_{\ell m \omega}^{\infty}}{\dd r_\star}-\tilde X_{\ell m \omega}^{\infty}\frac{\dd\tilde X_{\ell m \omega}^{r_+}}{\dd r_\star}=2i\omega \tilde A_{\ell m \omega}^{\rm in}
\label{eq:Wronskian_SN}
\end{equation}
is the (constant) Wronskian.

\begin{figure}[t]
    \centering
    \includegraphics[width=0.9\linewidth]{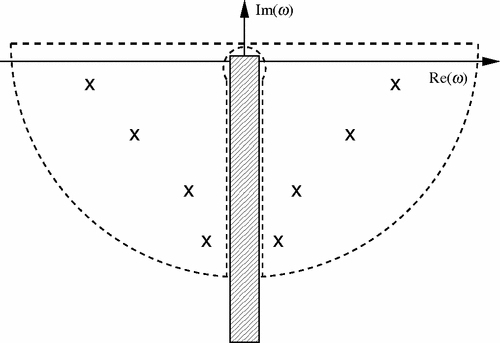}
    \caption{Integration contour to isolate the QNM contribution~\cite{Berti:2006wq}. The contour is closed in the lower semi-half including QNMs marked by crosses. The shaded are is the branch cut, whose contribution to the waveform is not discussed here.}
    \label{fig:poles}
\end{figure}

To find the spacetime perturbation in the time domain at future null infinity $\mathscr{I}^+$, which is related to the observed GW signal, we first compute the SN function in the time domain, i.e., the inverse Fourier transform of Eq.~\eqref{eq:SN_complete_solution}.
To define the asymptotic limit of Eq.~\eqref{eq:wf_SN_complete}, we introduce the null coordinates $u=t-r_\star$ and $v=t+r_\star$.
Future null infinity $\mathscr{I}^+$ corresponds to $v\to+\infty$ with constant $u$, i.e.
\begin{equation}
      \hat{\tilde X}_{\ell m}^p(u,v\to\infty)=\int_{\mathbb{R}}\dd \omega \ e^{-i\omega u}e^{-i\omega r_\star}\tilde X_{\ell m \omega}^p(r\to\infty) \,,
\label{eq:wf_SN_complete}
\end{equation}
with
\begin{equation}
\begin{aligned}
    &\tilde X_{\ell m \omega}^p(r\to+\infty)
    =\frac{e^{i \omega r_\star}}{2i\omega \tilde{A}_{\ell m\omega}^{\rm in}}\\
    & \int_{r_+}^\infty \dd r'\   X_{\ell m \omega}^{r_+}(r')\frac{W_{\ell m \omega}(r')}{r'^2(r'^2+a^2)^{1/2}}e^{- i\xi(r')}\ .
    \end{aligned}
    \label{eq:Xp_compute}
\end{equation}
Here we have used $X_{\ell m\omega}^{r_+}=\tilde X_{\ell m\omega}^{r_+}\sqrt{\gamma}$ and Eq.~\eqref{eq:W_def}.

In Eq.~\eqref{eq:wf_SN_complete}, the frequency is real, but we can isolate the QNM contribution by considering an analytic continuation of the variable $\omega$ to the complex plane, and by considering the integration contour shown in Fig.~\ref{fig:poles}.

At the QNM frequencies, computed here using the Leaver method (see Appendix~\ref{sec:Leaver_qnms}), the solutions of the homogeneous equation, $X_{\ell m\omega}^{r_+}$ and $X_{\ell m\omega}^{\infty}$, become proportional to each other~\cite{Leaver:1986gd}. Hence, the Wronskian in Eq.~\eqref{eq:Wronskian_SN} vanishes, and $\tilde A_{\ell m\omega}^{\rm in}=\tilde B_{\ell m\omega}^{\rm out}=0$: QNMs are simple poles of $\hat{\tilde X}^p_{\ell m}$.
If we neglect the tail contributions and the early response~\cite{Leaver:1986gd}, only the poles contribute to the integral:
\begin{align}
    &\hat{\tilde X}_{\ell m}^p(u,v\to\infty)=-\sum_{n}2\pi i\nonumber\\
    &\times\underset{\omega\to\omega_q}{\mbox{Res}}\int_{\mathbb{R}}\dd \omega \ e^{-i\omega t} \tilde X_{\ell m \omega}^p(r\to+\infty)\nonumber\\
    &=\sum_n \tilde C_q^{\rm SN} e^{-i\omega_q u}\,,
    \label{eq:wf_ringdown_TD}
\end{align}
where we have used $q=(\ell,m,n)$ for brevity. We have also defined the excitation coefficients for the rescaled SN function, 
\begin{equation}
    \tilde C_q^{\rm SN}=-\frac{\pi}{\omega_q\tilde\alpha_q}\int_{\mathbb{R}}\dd r' \  X_q^{r_+}(r')
     \frac{W_q(r')e^{- i\xi(r')}}{r'^2(r'^2+a^2)^{1/2}}\,,
     \label{eq:Cq_def}
\end{equation}
and we have used the Taylor expansion of $\tilde A_{\ell m\omega}^{\rm in}$ around $\omega=\omega_q$, i.e.
\begin{equation}
    \tilde A_{\ell m \omega}^{\rm in}\simeq \tilde\alpha_q (\omega-\omega_q) \qquad \tilde \alpha_q=\frac{\dd \tilde A_q^{\rm in}}{\dd\omega }\bigg|_{\omega=\omega_q} \ .
\end{equation}
In the excitation coefficients of Eq.~\eqref{eq:Cq_def}  we can distinguish two contributions: a source-independent factor $\pi/(\omega_q \tilde\alpha_q)$ and a source-dependent integral.
The source-independent \emph{excitation factor}~\cite{Berti:2006wq}, computed in Appendix~\ref{app:Teukolsky_amplitudes}, is commonly denoted as
\begin{equation}
    B_q=\frac{\tilde A_q^{\rm out}}{2\omega_q \tilde \alpha_{q}} \ ,\label{eq:excitation_factors}
\end{equation}
and it is independent of the normalization of the solution of the homogeneous equation.
We can then rewrite the source-dependent excitation coefficients in the compact form~\cite{Zhang:2013ksa} 
\begin{equation}
    \tilde C_q^{\rm SN}=I_q B_q\ , \label{eq:excitation_coefficients}
\end{equation}
where
\begin{equation}
  I_q=-2\pi\frac{\sqrt{{c_{0}}}}{A_q^{\rm out}}\int_{r_+}^\infty \dd r'\  X_q^{r_{+}}(r')\frac{W_q(r')}{r'^2(r'^2+a^2)^{1/2}}e^{- i\xi(r')} \;, \label{eq:Iq}
\end{equation}
$A_q^{\rm out}=\sqrt{c_{0}}\ \tilde A_q^{\rm out}$, the outgoing amplitude of the non-rescaled SN function, is computed in Appendix~\ref{app:Teukolsky_amplitudes}, and we recall that, for simplicity, we omit the dependence of $c_0$ on the multi-index $q=(\ell m n)$.
The integral $I_q$ is, in general, divergent for a point-particle source. 
This divergence is removed when the full waveform is reconstructed and causality is consistently taken into account~\cite{DeAmicis:2025xuh}. 
The divergent part gives rise to a redshifted mode, whose contribution to the waveform vanishes at $\mathscr{I}^+$.

Since we are only interested in QNM excitation, we adopt a regularization procedure (see Appendix~\ref{app:Iq_reg}) to remove the divergence in $I_q$ and we ignore the contribution due to the redshifted mode.
From the regularized integral $\mathcal I_q$ we can then compute the excitation coefficients $\tilde C_q^{\rm SN}$ for the rescaled SN function, defined in Eq.~\eqref{eq:excitation_coefficients}, 
as
\begin{equation}
    \tilde C_q^{\rm SN}=\  B_q \mathcal I_q \; . 
    \label{eq:C_q_fin}
\end{equation}
\subsection{The excitation coefficients of the gravitational strain}\label{sec:hexcitationcoefficients}
By Fourier-transforming and expanding in harmonics the gravitational strain, from Eqs.~\eqref{eq:psi4h} and \eqref{eq:expansion_psi} we have:
\begin{equation}
    h(t,r,\theta,\phi)=-2\sum_{\ell m}
   \frac{\rho^4}{\sqrt{2\pi}} \int d\omega 
   \frac{e^{-i\omega t}}{\omega^2}\,_2S^{a\omega}_{\ell m}(\theta,\phi)
    R_{\ell m\omega}(r)\,.
\end{equation}
In the limit $r\to\infty$ we have $\rho\simeq 1/r$ and (see e.g.~\cite{Nakamura:1987zz,Watarai:2024huy})
\begin{equation}
 R_{\ell m\omega}\simeq-\frac{4\omega^2r^3}{c_{0}}X_{\ell m\omega}\,.
\end{equation}
From Eqs.~\eqref{eq:gamma_ci_def}, \eqref{eq:c_0}, and \eqref{eq:tilde_chi_def} we also have ${\tilde X}_{\ell m\omega}\simeq X_{\ell m\omega}/\sqrt{c_{0}}$. 
Thus, in the limit $t\to\infty$, $h\simeq\sum_{\ell m}h_{\ell m}$ with
\begin{align}
    h_{\ell m}&=\frac{8}{\sqrt{2\pi}}\frac{1}{r}\int d\omega e^{-i\omega t}\,_{-2}S^{a\omega}_{\ell m}(\theta,\phi)\frac{{\tilde X}_{\ell m\omega}}{\sqrt{c_0}}\nonumber\\
    &=\frac{8}{\sqrt{2\pi}}\frac{1}{r}\sum_n {\tilde C}^{\rm SN}_{\ell mn}e^{-i\omega_{\ell mn}u}\frac{\,_{-2}S^{a\omega}_{\ell m}(\theta,\phi)}{\sqrt{c_0}}\,,
\end{align}
where we have performed the same integration as in Eq.~\eqref{eq:wf_ringdown_TD}, using the contour shown in Fig.~\ref{fig:poles}.
Finally, by comparison with Eq.~\eqref{eq:mode_expansion}, we find that the excitation coefficients of the gravitational strain are
\begin{equation}
    C_q=\frac{8}{\sqrt{2\pi c_{0}}}\tilde C^{\rm SN}_q\,.\label{eq:Cfin}
\end{equation}

\section{The source term for a particle on a critical plunge geodesic}\label{sec:sourceterm}

We compute the excitation coefficients assuming that the source of the Teukolsky equation describes a plunging particle following a critical plunge geodesic~\cite{Mummery:2022ana,Dyson:2023fws}. Following the procedure in Sec.~\ref{sec:theorysetup}, we start from the Teukolsky framework and then map all relevant quantities to the SN approach.

The source term of the radial Teukolsky equation \eqref{eq:R_teukolsky} is given by~\cite{Breuer:1975book,Spiers:2024src,Nakamura:1987zz}
\begin{equation}
\label{eq:T_lmw}
    T_{\ell m\omega} = \sqrt{\frac{8}{\pi}}\int \frac{S(\theta)}{\rho^{5} \overline{\rho}} \left(B'_2+B'^*_2\right) \mathrm{e}^{-i(\omega t - m\phi)}\dd\Omega \dd t\ ,
\end{equation}
where the overbar denotes complex conjugation,
\begin{align}
B'_2 =& -\frac{1}{2} \rho^8 \Bar{\rho} L_{-1} \left[ \rho^{-4} L_0 \left[ \rho^{-2} \Bar{\rho}^{-1} T_{nn} \right] \right]\nonumber\\
& -\frac{1}{2\sqrt{2}} \rho^8 \Bar{\rho} \Delta^2 L_{-1} \left[ \rho^{-4} \Bar{\rho}^2 J_+ \left[ \rho^{-2} \Bar{\rho}^{-2} \Delta^{-1} T_{\bar{m}n} \right] \right]\;,\\
B'^*_2 =&-\frac{\Delta^2 \rho^8 \bar{\rho}}{2 \sqrt{2}} J_+\left[\bar{\rho }^2 \Delta^{-1}\rho ^{-4}L_{-1}^\dagger\left[T_{\bar{m}n}\rho^{-2} \bar{\rho }^{-2}\right]\right]\nonumber\\
&-\frac{1}{4} \Delta ^2 \rho^8 \bar{\rho }J_+\left[\rho^{-4}J_+\left[T_{\bar{m}\bar{m}}\bar{\rho }\rho^{-2}\right]\right]\;,
\end{align}
and 
\begin{align}
    L_s&=\partial_\theta+m \csc\theta-a\omega\sin\theta+s\cot\theta\; ,\\
   J_+&=\partial_r+i \frac{K}{\Delta} \ .
\end{align}
This definition of $T_{\ell m\omega}$ differs from that found, e.g., in Ref.~\cite{Hughes:1999bq}, due to the different normalization of the spheroidal functions, which for brevity we denote as $\,_{-2}S^{a\omega}_{\ell m}(\theta,\phi)=S(\theta)e^{im\phi}$: see Eq.~\eqref{eq:S_norm_cond} in Appendix~\ref{sec:Leaver_qnms}.

The coefficients $T_{nn}$, $T_{\bar{m}n}$ and $T_{\bar{m}\bar{m}}$ are obtained by projecting the stress–energy tensor onto the Kinnersley tetrad~\cite{Kinnersley:1969zza}, whose vector basis in Boyer–Lindquist coordinates reads
\begin{subequations}\label{eq:tetrad}
\begin{eqnarray}
    l^\mu&=&\frac{1}{\Delta}\left(r^2+a^2,\Delta,0,a\right)\\
    n^\mu&=&\frac{1}{2\Sigma}\left(r^2+a^2,-\Delta,0,a\right)\\
    m^\mu&=&\frac{1}{\sqrt{2}}\bar{\rho}\left(ia \sin\theta,0,1,i\csc\theta\right)\\
    \overline{m}^\mu&=&\frac{1}{\sqrt{2}}\rho\left(-ia \sin\theta,0,1,-i\csc\theta\right).
    \end{eqnarray}
\end{subequations}
For a point particle of mass $m_p$, the stress–energy tensor $T^{\mu\nu}$ is
\begin{equation}
    T^{\mu\nu}=\frac{m_p}{\Sigma \sin \theta\dot t}\dot x^\mu\dot x^\nu \delta(r-r(t))\delta(\theta-\theta(t))\delta(\phi-\phi(t))\ ,
\end{equation}
where $\Sigma=1/(\rho\bar\rho)=r^2+a^2\cos^2\theta$, the equations of motion $\dot x^\mu=\dd x^\mu/\dd \tau$ are given explicitly in Sec.~\ref{sec:geodesics}, and $r(t)\ ,\theta(t)$, and $\phi(t)$ denote the orbital elements as functions of the coordinate time $t$.
The tetrad projections can be written as
\begin{equation}
    T_{ab}=m_p\frac{C_{ab}}{\sin\theta} \delta(r-r(t))\delta(\theta-\theta(t))\delta(\phi-\phi(t))\ ,
\end{equation}
where the coefficients $C_{nn}, \ C_{\bar{m}n}$ and $C_{\bar{m}\bar{m}}$ are
\begin{subequations}
\begin{align}
    C_{nn}&=\frac{(\dot{r} \Sigma+P)^2}{4 \dot t \Sigma^3}\ ,\\
    C_{\bar{m}n}&= -\frac{\rho (\dot r \Sigma +P) }{2 \sqrt{2} \dot t \Sigma ^2}\left[\dot \theta \Sigma+i \sin\theta\left(a \EN-\ANG \csc ^2\theta\right)\right] \ ,\\ 
    C_{\bar{m}\bar{m}}&= \frac{\rho^2}{2 \dot t \Sigma} \left[\dot \theta \Sigma +i \sin \theta  \left(a \EN-\ANG \csc ^2\theta \right)\right]^2\ .
\end{align}
\label{eq:C_T_tetrad}
\end{subequations}
Here we have defined the function $P= \EN (a^2+r^2)-a \ANG$, where $\EN$ and $\ANG$ are the energy and angular of momentum (per unit mass) of the particle at infinity.

The source term in Eq.~\eqref{eq:T_lmw} can be simplified using the identity
\begin{equation}
    \int_0^\pi \dd\theta \sin \theta \ v(\theta) L_{s}[w(\theta)]=-\int_0^\pi \dd \theta \sin\theta \ L_{1-s}^\dagger[v(\theta)] w(\theta)\ ,
\end{equation}
where $L^\dagger_{s}=\partial_\theta-m\csc\theta+a\omega\sin\theta+s\cot\theta$, yielding
\begin{widetext}
\begin{align}
    T_{\ell m\omega}=
    &\frac{4m_p}{\sqrt{2\pi }}\int\dd t\ e^{i [\omega t - m \phi(t)] } \Bigg\{\frac{\Delta C_{\bar{m}n}\delta(r-r(t)) }{2 \sqrt{2} {\bar{\rho}}^2 \rho^2} L^\dagger_2\left[S \rho ^3 \left(\frac{2 \bar{\rho}\bar{\rho}_{,r}}{\rho^4}-\frac{4 {\bar{\rho}}^2 \rho_{,r}}
    {\rho^5}\right)\right]-\frac{1}{4}
   S \Delta^2 \rho^3 J_+\left[\rho ^{-4}J_+\left[\frac{C_{\bar{m}\bar{m}}\delta(r-r(t))\bar{\rho}}{\rho^2}\right]\right]\nonumber\\
   &+\frac{\Delta^2 \bar{\rho}}{\sqrt{2} \rho^2}L^\dagger_2\left[S \rho \bar{\rho}\right] J_+\left[\frac{C_{\bar{m}n}}{\Delta \rho^2 {\bar{\rho}}^2}\delta(r-r(t))\right]
   -\frac{ L_1^\dagger\left[\rho^{-4}L^\dagger_2\left[S \rho^3\right]\right]}{2 \bar{\rho}\rho^2}C_{nn}\delta(r-r(t))\Bigg\} \; . \label{eq:T_fin}
\end{align}
\end{widetext}
Here all functions are evaluated at $\theta=\theta(t)$, i.e., along the particle’s polar trajectory. 

From Eqs.~\eqref{eq:W_T_link} and~\eqref{eq:W_def} we can get $W$ and $\mathcal{S}_{\ell m\omega}$~\cite{Nakamura:1987zz,Watarai:2024huy} (see Appendix~\ref{app:W_generic} for further details).

\subsection{Critical plunge geodesics}\label{sec:geodesics}
We now focus on the case of particles plunging from the innermost stable spherical orbit (ISSO),
building upon closed form analytical expressions of geodesic motion in terms of elliptic functions~\cite{Dyson:2023fws}.
Here we consider the general case of {\em inclined} plunging trajectories for future reference, but our numerical analysis will focus (for the time being) on equatorial orbits. 
\subsubsection{Geodesic Equations}\label{sec:geo_eqs}
Geodesic motion in the Kerr spacetime has four conserved quantities:
\begin{align}
    \EN &= -u^{\mu}g_{\mu\nu}\left(\frac{\partial}{\partial t}\right)^{\nu}\;,\label{eq:geo:energy} \\
    \ANG &=  u^{\mu}g_{\mu\nu}\left(\frac{\partial}{\partial \phi}\right)^{\nu}\;, \label{eq:geo:angular_momentum}\\ 
     Q &=  u^{\mu}\mathcal{K}_{\mu\nu}u^{\nu} - (\ANG-a\EN)^2\; , \label{eq:geo:Carter_constant}\\
     -1&=g_{\mu\nu}u^\mu u^\nu\label{eq:geo:mass_shell_condition}\; ,
\end{align}
where $Q$ is the Carter constant~\cite{Carter:1968rr}.
The quantity
\begin{equation}
	\mathcal{K}_{\mu\nu} = 2\Sigma l_{(\mu} n_{\nu)} + r^2 g_{\mu\nu}\; , \label{eq:Killing_tensor}
\end{equation}
is the so-called Killing-Yano tensor~\cite{Frolov:2017kze}, while $l_{\mu}$ and $n_{\nu}$ are the principal null vectors of the Kinnersley tetrad: see Eq.~\eqref{eq:tetrad}.
Using these constants of motion, the geodesic equations can be written as
\begin{align}
    &\begin{aligned}\label{eq:geo:radialeom}
        \left(\frac{\dd r}{\dd\lambda_{\rm M}}\right)^2 
        &= (\mathcal{E}(r^2+a^2)-a\mathcal{L})^2-\Delta(r^2+(a\mathcal{E}-\mathcal{L})^2+Q)\\
        &= (1-\EN^2)(r_1-r)(r_2-r)(r_3-r)(r-r_4) \\
        &= \mathcal{R}(r), 
    \end{aligned}\\
    &\begin{aligned}\label{eq:geo:polareom}
        \left(\frac{\dd z}{\dd\lambda_{\rm M}}\right)^2 &= Q-z^2[a^2(1-\EN^2)(1-z^2) + \mathcal{L}^2+Q]\\
        &= (z^2-z_1^2)(a^2(1-\EN^2) z^2-z_2^2)\\
        &= Z(z),
    \end{aligned}\\
    &\;\;\begin{aligned}\label{eq:geo:timeeom}
        \frac{\dd t}{\dd\lambda_{\rm M}} &= \frac{(r^2+a^2)}{\Delta}(\mathcal{E}(r^2+a^2)-a\mathcal{L})-a^2\mathcal{E}(1-z^2)+a\mathcal{L}, 
    \end{aligned}\\
    &\;\;\begin{aligned}\label{eq:geo:azimuthaleom}
        \frac{\dd \phi}{\dd \lambda_{\rm M}} &= \frac{a}{\Delta}(\mathcal{E}(r^2+a^2)-a\mathcal{L})+\frac{\mathcal{L}}{1-z^2}-a\mathcal{E}.
    \end{aligned}
\end{align}
Here $z=\cos\theta$, and $\lambda_{\rm M}$ is the Mino time~\cite{Mino:2003yg}, defined as 
\begin{equation}
    \dd\tau = \Sigma \dd\lambda_{\rm M}\; .
    \label{eq:Mino_time}
\end{equation}
The quantities $r_1, r_2, r_3,$ and $r_4$ are the roots of the radial potential $\mathcal{R}(r)$, while
\begin{equation}
z_1=\frac{(1+Z_-)^{1/2}}{\sqrt{2}}\ ,\quad 
z_2=\frac{a(1-\EN^2)^{1/2}}{\sqrt{2}}(1+Z_+)^{1/2}\ ,
\end{equation}
with
\begin{equation}
Z_{\pm}=\frac{\mathcal{L}^2+Q}{a^2(1-\EN^2)}\pm\sqrt{\left[1 + \frac{\mathcal{L}^2+Q}{a^2(1-\EN^2)}\right]^2-\frac{4Q}{a^2(1-\EN^2)}}
\end{equation}
are the roots of the polar potential, i.e., $Z(z)=0$.
Using the parametrization~\eqref{eq:Mino_time}, the equations of motion fully decouple and can be solved in closed form in terms of elliptic integrals~\cite{Dyson:2023fws,vandeMeent:2019cam}. 

Plunging geodesics are bounded between two roots $r_i,r_j$ of $\mathcal{R}(r)=0$ such that
\begin{equation}
    r_i< r_-< r_+ < r_j \; .\label{eq:plunge_conds}
\end{equation}
If the radial potential $\mathcal{R}(r)$ associated with a plunging geodesic has no vanishing roots and all its roots are simple (i.e., of multiplicity one), 
the corresponding trajectory has no radial turning points. The particle therefore crosses both the outer and inner horizons ($r_+$ and $r_-$, respectively).

In the special case of an equatorial trajectory ($Q = 0$), as those considered in Ref.~\cite{Mummery:2022ana}, the point $r = 0$ itself is a root of the radial equation. It then acts as the inner turning point, and the orbit terminates there within a finite amount of proper time. Conversely, if $\mathcal{R}(r)$ admits a multiple root, the corresponding radius is approached asymptotically, and the particle 
reaches it only after an infinite amount of proper time.

Two roots fulfilling Eq.~\eqref{eq:plunge_conds} can always be found when $\EN<1$ and $Q\ge 0$~\cite{Dyson:2023fws}. Additional plunging orbits can exist for $\EN>1$ and $Q<0$, corresponding to trajectories plunging from infinity while oscillating poloidally around $z_0\ne 0$.

\subsubsection{Plunging Orbits from the Innermost Stable Spherical Orbit}\label{subsubsecISSO}
The ISSO radius $r_{\rm I}$ is a triple root of $\mathcal{R}(r)$: $\mathcal{R}(r_{\rm I})=\mathcal{R}'(r_{\rm I})=\mathcal{R}''(r_{\rm I})=0$. 
When $r_1=r_2=r_3=r_{\rm I}$, Eq.~\eqref{eq:geo:radialeom} reduces to 
\begin{equation}
\left(\frac{\dd r}{\dd\lambda_{\rm M}}\right)^2=\left(1-\EN^2\right)(r_{\rm I}-r)^3(r-r_4) \; ,\label{eq:geo:ISSO:radialeom}
\end{equation}
where 
\begin{equation}
    r_4=\frac{a^2Q}{\left(1-\mathcal{E}^2\right) r_{\rm I}^3}\; , \quad r_4< r_-<r_+<r_{\rm I} \; .
\end{equation}
Using the explicit form of $\mathcal{R}$ in Eq.~\eqref{eq:geo:radialeom}, $r_{\rm I}$ can be determined in terms of $a$ and $z_1$~\cite{Stein:2019buj,Ng:2025maa}.
The constants of motion ${\cal E},{\cal L},Q$ at the ISSO are
\begin{align}
    Q &= r_{\rm I}^{\frac{5}{2}}\frac{(\sqrt{(r_{\rm I}-r_{+})(r_{\rm I}-r_{-})}-2\sqrt{r_{\rm I}})^2-4a^2}{4a^2(\sqrt{(r_{\rm I}-r_{+})(r_{\rm I}-r_{-})}+\sqrt{r_{\rm I}}-r_{\rm I}^{\frac{3}{2}})}\; ,\label{eq:geo:ISSO:carter}\\
    \EN &= \frac{\sqrt{a^2Q-2r_{\rm I}^3+3r_{\rm I}^4}}{\sqrt{3}r_{\rm I}^2}\label{eq:geo:ISSO:en}\; ,\\
    \ANG &= \epsilon_\pm\frac{\sqrt{3a^2Q-a^2r_{\rm I}^2-Qr_{\rm I}^2+3r_{\rm I}^4+a^2 r_{\rm I}^2\EN^2-3r_{\rm I}^4\EN^2}}{r_{\rm I}}\label{eq:geo:ISSO:ANG} \; .
\end{align}
The parameter $\epsilon_\pm$ in Eq.~\eqref{eq:geo:ISSO:ANG} 
depends on the ratio $r_{\rm I}/\mathcal{L}_{\rm root}$:
\begin{equation}
 \epsilon_\pm=\begin{cases}
     + , & \text{if } r_{\rm I} \le \ANG_{\rm root},\\
     - , & \text{if } r_{\rm I} > \ANG_{\rm root},
 \end{cases}
\end{equation}
where $\ANG_{\rm root}$ is the real root closest to $r=6$ 
on the right-hand side of Eq.~\eqref{eq:geo:ISSO:ANG}.

The equations for the radial and polar motion can be solved in 
terms of elliptic functions:
\begin{align}
     r(\lambda_{\rm M}) &= \frac{r_{\rm I}(r_{\rm I}-r_4)^2(1-\EN^2)\lambda_{\rm M}^2+4r_4}{(r_{\rm I}-r_4)^2(1-\EN^2)\lambda_{\rm M}^2+4}\; , \label{eq:geo:ISSO:rsoleom}\\
     z(\lambda_{\rm M})&= z_1\sin\left[\mbox{am}(z_2\lambda_{\rm M}|k_z^2)\right]\; ,\label{eq:geo:chisoleom}
\end{align}
where $\mbox{am}(\cdot,\cdot)$ is the Jacobi amplitude and 
\begin{equation}
    k_z^2=a^2\left(1-\EN^2\right)\frac{z^2_1}{z^2_2}\; .
\end{equation}
Finding solutions for the azimuthal and temporal equations is more 
involved since their right-hand sides depend on both 
$r$ and $z$. However, 
Eqs.~\eqref{eq:geo:timeeom}--\eqref{eq:geo:azimuthaleom} 
admit solutions of the form 
\begin{align}
    \phi(\lambda_{\rm M})&=\phi_r(r(\lambda_{\rm M}))+\phi_z(z(\lambda_{\rm M}))-a\EN\lambda_{\rm M}\ ,\label{eq:geo:solphi}\\
t(\lambda_{\rm M})&=t_r(r(\lambda_{\rm M}))+t_z(z(\lambda_{\rm M}))+a\ANG \lambda_{\rm M}\label{eq:geo:solt}\ ,
\end{align}
where
\begin{align}
\phi_z(\lambda_{\rm M})&=\frac{\mathcal{L} \Pi \left(z_1^2;\text{am}\left(\lambda_{\rm M} z_2|k_z^2\right)|k_z^2\right)}{z_2}\ ,\\
t_z(\lambda_{\rm M})&= - \frac{\mathcal{E}}{1-\mathcal{E}^2}\, z_2\, E\left(\text{am}\left(\lambda_{\rm M} z_2|k_z^2\right)|k_z^2\right)\ ,
\end{align}
and
\begin{widetext}
\begin{align}
\phi_r(\lambda_{\rm M})=&\frac{a \lambda_{\rm M}  \left((a^2+r_{\rm I})^2 \mathcal{E}-a \mathcal{L}\right)}{\left(r_{\rm I}-r_-\right) \left(r_{\rm I}-r_+\right)}+\frac{a \left((a^2+r_-^2) \mathcal{E}-a \mathcal{L}\right)}{2 \left(r_+-r_-\right)
   \left(r_{\rm I}-r_-\right)^{3/2} \sqrt{\left(r_--r_4\right) \left(1-\mathcal{E}^2\right)}} \nonumber\\
&\times\log\left(\frac{\left(2-\lambda_{\rm M}  \left(r_4-r_{\rm I}\right)
   \sqrt{\frac{\left(\mathcal{E}^2-1\right)
   \left(r_{\rm I}-r_-\right)}{r_4-r_-}}\right)^{2}}{\left(2+\lambda_{\rm M}  \left(r_4-r_{\rm I}\right)
   \sqrt{\frac{\left(\mathcal{E}^2-1\right)
   \left(r_{\rm I}-r_-\right)}{r_4-r_-}}\right)^{2}}\right)+(r_- \leftrightarrow r_+)\ ,\\
t_r(\lambda_{\rm M})=&\frac{\lambda_{\rm M}  \left(a^2+r_{\rm I}^2\right) \left(a^2 \mathcal{E}-a \mathcal{L}+r_{\rm I}^2\mathcal{E}\right)}{\left(-r_-+r_{\rm I}\right) \left(-r_++r_{\rm I}\right)}+\frac{2 \lambda_{\rm M} 
\left(r_4-r_{\rm I}\right)^{2} \mathcal{E}}{4+\lambda_{\rm M}^{2} \left(r_4-r_{\rm I}\right)^{2}
\left(1-\mathcal{E}^2\right)}-\frac{2 \left(r_-+r_+\right)+3 r_{\rm I}+r_4}{\sqrt{1-\mathcal{E}^2}}\mathcal{E}\nonumber\\
&\times\tan ^{-1}\left[\frac{1}{2} \lambda_{\rm M}  \left(-r_4+r_{\rm I}\right)
   \sqrt{1-\mathcal{E}^2}\right]+\frac{\left(a^2+r_-^2\right) \left(\mathcal{E} \left(a^2+r_-^2\right)-a
\mathcal{L}\right)}{2 \left(r_+-r_-\right)\left(-r_-+r_{\rm I}\right)^{3/2} \sqrt{\left(r_--r_4\right)\left(1-\mathcal{E}^2\right)}}\nonumber\\
&\times\log \left(\frac{\left(2-\lambda_{\rm M}  \left(r_4-r_{\rm I}\right)
\sqrt{\frac{\left(-r_-+r_{\rm I}\right)
\left(\mathcal{E}^2-1\right)}{r_4-r_-}}\right)^{2}}{\left(2+\lambda_{\rm M} \left(r_4-r_{\rm I}\right) \sqrt{\frac{\left(-r_-+r_{\rm I}\right)\left(\mathcal{E}^2-1\right)}{r_4-r_-}}\right)^{2}}\right)+(r_- \leftrightarrow r_+)\ .   
\end{align}
\end{widetext}
In the case of an equatorial plunge, $r_{\rm I}$ coincides with the ISCO, and we can evaluate Eqs.~\eqref{eq:geo:ISSO:rsoleom},~\eqref{eq:geo:chisoleom},~\eqref{eq:geo:solphi} and \eqref{eq:geo:solt} with $Q=r_4=z_1=0$ to describe the particle trajectory.

\section{Results}\label{sec:results}

The computation of the excitation coefficients~\eqref{eq:C_q_fin} of particles in critical plunging equatorial orbits requires specifying the values of $\EN$, $\ANG$ corresponding to these geodesics, and the particle’s initial position at the ISCO, $r_p = r_{\rm I}$. These quantities depend on three parameters: the BH mass $M$, the spin $a$, and the particle mass $m_p$. 

Since the dependence on $m_p$ can be factored out ($C_q \propto m_p$), in what follows we shall focus on the dependence of the excitation coefficients on  the dimensionless spin parameter $a/M$. 
%
\subsection{Comparison between numerical waveform and quasinormal modes superposition}\label{sec:comparison}
In order to check the validity of our results and to assess the accuracy of the QNM expansion, we can compare the QNM sum in terms of the excitation coefficients of Eq.~\eqref{eq:mode_expansion} with the numerical solution of the SN equation found by Green's function methods as in Eq.~\eqref{eq:wf_SN_complete}, and then Fourier-transformed back to the time domain.

\begin{figure}[t]
    \centering
\includegraphics[width=0.98\linewidth]{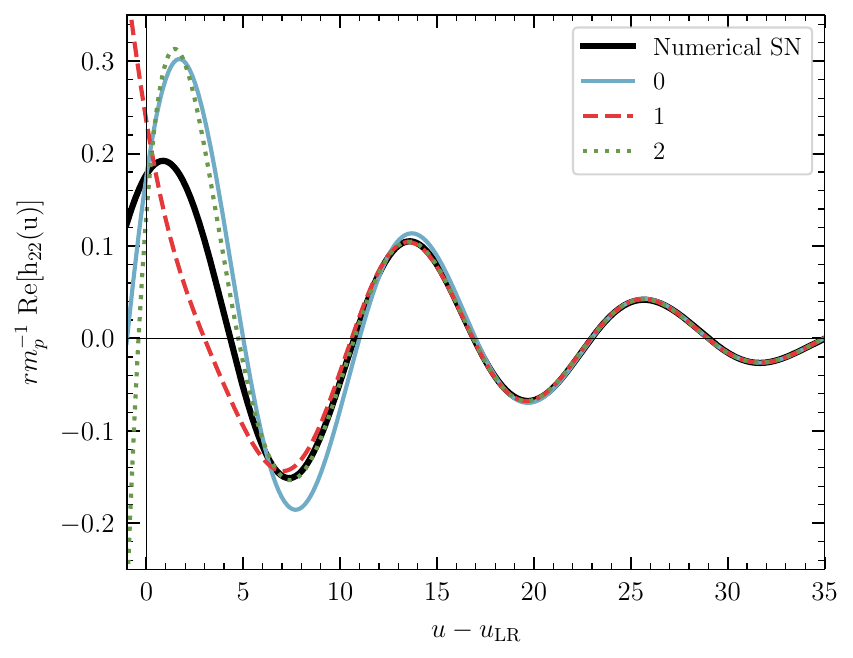}
    \caption{\small{Comparison between the real part of the full SN function ~\eqref{eq:wf_SN_complete} (thick solid  black line) and of the mode superposition~\eqref{eq:mode_expansion}   including QNMs up to the overtone number $n$ shown in the legend (thin colored lines). 
    We assume $a/M=0.68$ and $\ell =m=2$.  Here, $u_{\rm LR}$ is the position of the light ring  in terms of the null coordinate $u=t-r_\star$. } }
    \label{fig:comparison}
\end{figure}

In Fig.~\ref{fig:comparison} we show a comparison between these two quantities for the dominant $\ell=m=2$ term in the multipolar expansion $h=\sum_{\ell m}h_{\ell m}$. We consider an astrophysically motivated value of the spin $a/M=0.68$
and different choices of the maximum overtone number in Eq.~\eqref{eq:mode_expansion}.
As expected, including higher overtones improves the agreement between the QNM approximation and the full numerical solution. Qualitatively, a QNM approximation including only the fundamental mode reproduces the full SN function for $ u - u_{\rm LR} \gtrsim 15\,M $. 
Adding two overtones improves the agreement, extending the region in which the solutions overlap down to $u - u_{\rm LR} \gtrsim 10\,M $ ($5\,M $) when we include up to the first (second) overtone.
This behavior is consistent with theoretical expectations, and provides a numerical validation of our calculation of the excitation coefficients.
%

\subsection{Excitation coefficients as functions of the spin}
\label{sec:excitation_coefficients_a_fundamental}
%

We can now explore in more detail how the excitation coefficients $C_q$ vary with the BH spin, and compare the excitation coefficients corresponding to different modes. The contribution of the modes to the detected strain depends on the inclination of the source with respect of the observer through the spin-weighted spheroidal harmonics appearing in the mode expansion~\eqref{eq:mode_expansion} (see e.g.~\cite{Watarai:2024huy}), but the excitation coefficients $C_q$ are still useful to quantify the amplitude of different modes.

\subsubsection{Excitation coefficients for $\ell=m=2$}
\begin{figure}[t]
    \centering
    \includegraphics[width=0.98\linewidth]{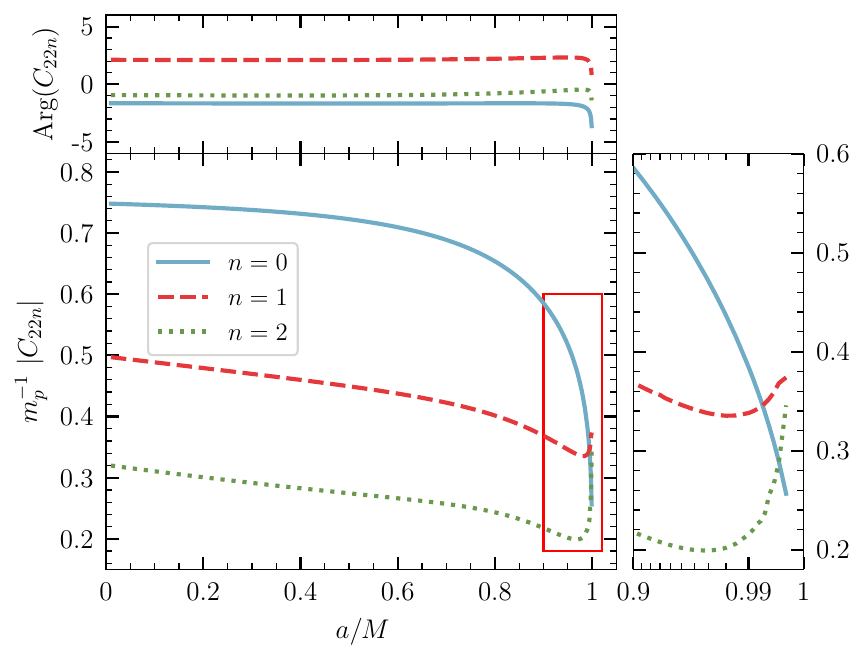}
    \caption{
    Excitation coefficients for $\ell = m = 2$ and different 
    overtone numbers $n$ as functions of $a/M$. The solid blue, dashed red and dotted green lines refer to the fundamental mode ($n=0$), the first overtone ($n=1$) and the second overtone ($n=2$), respectively. The main (top) panel shows the absolute value     (phase) of the excitation coefficients. The right panel is a zoomed-in view of the absolute value in the region $a/M > 0.9$, highlighted by a red rectangle in the main panel. This close-up reveals that for rapidly rotating BHs ($a/M \gtrsim 0.994$) the fundamental mode does not correspond to the largest value of $|C_q|$.     }
    \label{fig:exc_coeff_main}
\end{figure}
In Fig.~\ref{fig:exc_coeff_main} we display the absolute value and the phase of $C_q$ for $\ell = m = 2$ QNMs with $n = 0,\,1,\,2$ as functions of $a/M$. We observe that the absolute value of the excitation coefficient $|C_q|$ for the fundamental mode ($n=0$) decreases monotonically with $a/M$, while those of the overtones ($n>0$) decrease up to $a/M \simeq 0.99$, reach a minimum, and then grow again at higher spins. 
The fundamental QNM is the most strongly excited for $0 \le a/M \lesssim 0.994$, and $|C_{221}|>|C_{222}|$ at any spin. However, as highlighted in the right panel of Fig.~\ref{fig:exc_coeff_main}, which zooms into the high-spin region, $|C_q|$ for the first (second) overtone becomes larger than $|C_q|$ for the fundamental mode when $a/M\gtrsim0.994$ ($a/M\gtrsim0.998$).

Extrapolating these results to less asymmetric binaries suggests that overtones may leave a significant imprint in the ringdown of rapidly rotating BHs formed in mergers of comparable-mass or intermediate mass-ratio systems. This possibility is interesting in view of the expected high-SNR observations of SMBH mergers by LISA, where highly spinning remnants may be common~\cite{Berti:2008af,Barausse:2012fy}. 

The phase of the excitation coefficients shows a milder dependence on the spin. For all modes, ${\rm Arg}(C_q)$ remains nearly constant for $a/M \lesssim 0.9$ at the level of a few percent. For $a/M \gtrsim 0.99$, however, the phase changes abruptly and decreases monotonically for all of the values of $n$ we considered.

\begin{figure}[t]
    \centering
    \includegraphics[width=0.98\linewidth]{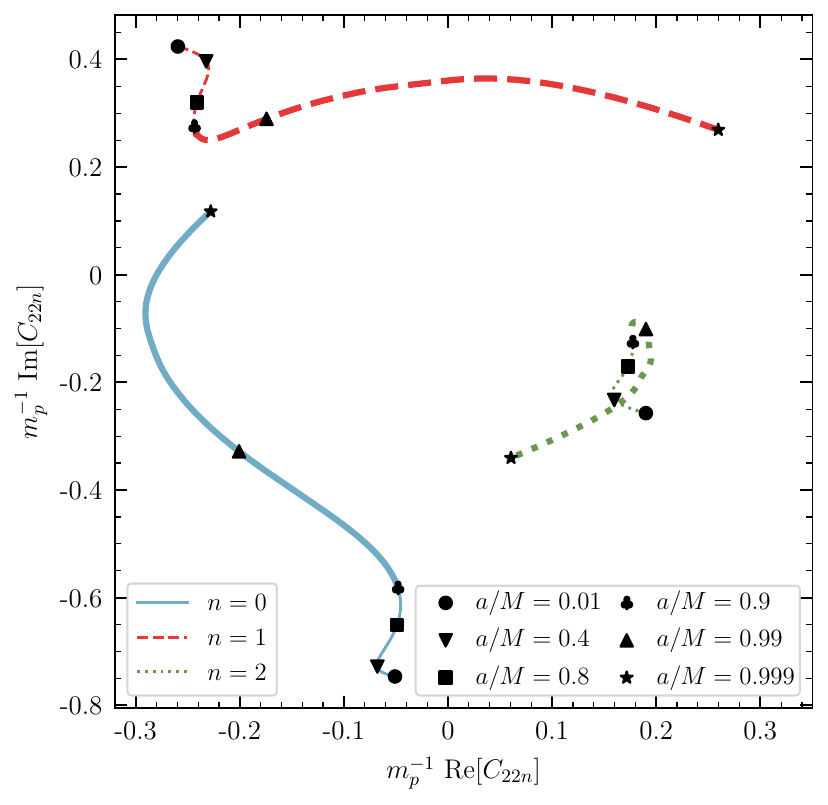}
    \caption{Excitation coefficient trajectories, parametrized by $a/M$, for $\ell = m = 2$ and different overtone
    numbers. The solid blue, dashed red and dotted green lines refer to the fundamental mode ($n=0$), the first overtone ($n=1$) and the second overtone ($n=2$), respectively. 
    Thick lines highlight the near-extremal regime $a/M > 0.9$, while markers indicate selected spin values, as identified in the legend.} 
    \label{fig:exc_coeff_re_vs_im}
\end{figure}

In Fig.~\ref{fig:exc_coeff_re_vs_im} we show the real and imaginary parts of the excitation coefficients in the complex plane.
Thicker lines correspond to near-extremal BHs ($a/M > 0.9$). These trajectories clearly illustrate that the QNM excitation  is considerably more sensitive to spin variations in the high-spin regime, where small variations of the spin cause variations of $C_q$ over much larger ranges relative to the subextremal interval ($0 \le a/M < 0.9$).

\subsubsection{Excitation coefficients for higher harmonics}
\begin{figure}[t]
    \centering
    \includegraphics[width=0.98\linewidth]{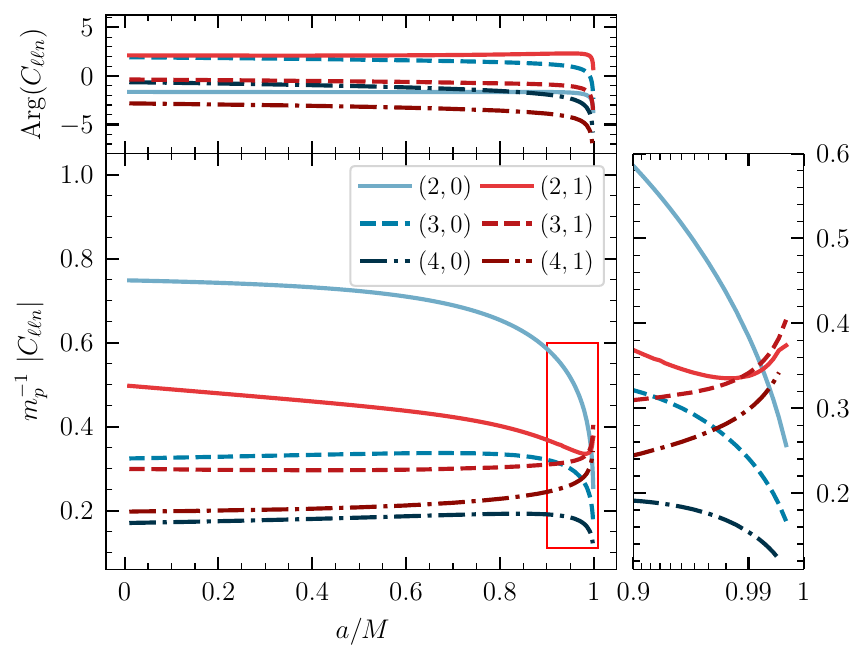}
   \caption{
     Same as Figure~\ref{fig:exc_coeff_main}, but for $\ell=m=2,\,3,\,4$ and $n=0,\,1$. The legend shows the pair $(\ell=m,\,n)$. Lines in different shades of blue (red) correspond to $n=0$ ($n=1$). We observe that $|C_{220}|$ is the largest for $a/M\lesssim0.994$, while the values of $|C_q|$ for all modes with $n=1$ become larger than $|C_{220}|$ close to extremality.} 
    \label{fig:exc_coeff_HM_01}
\end{figure}

\begin{figure}[t]
    \centering
    \includegraphics[width=0.98
\linewidth]{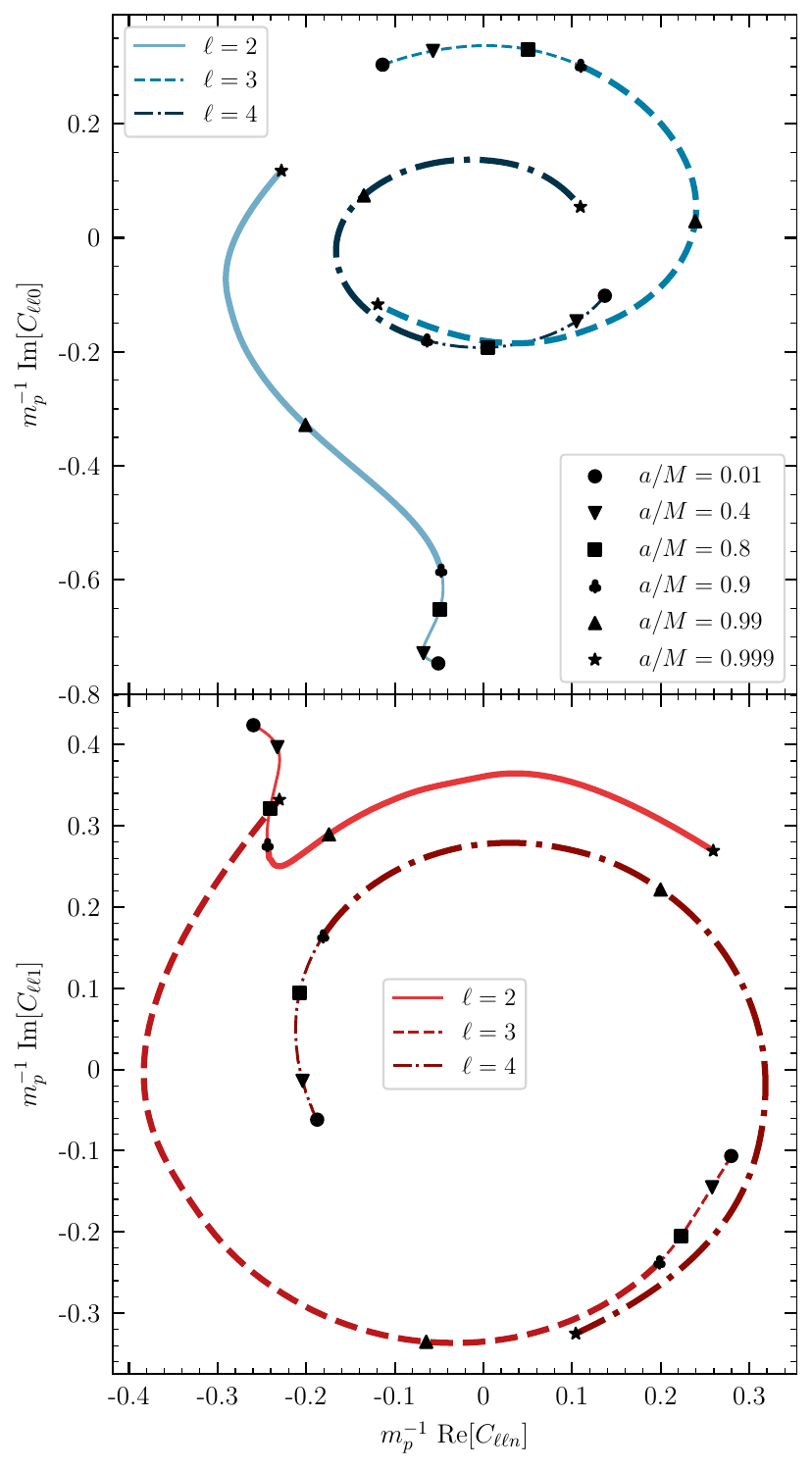}
   \caption{
   Same as Fig.~\ref{fig:exc_coeff_re_vs_im}, but for the excitation coefficients of QNMs with $\ell=m=2,\,3,\,4$. The upper and lower panels refer to $n=0$ and $n=1$, respectively.
   }
    \label{fig:exc_coeff_HM}
\end{figure}

In Fig.~\ref{fig:exc_coeff_HM_01} we show the excitation coefficients for QNMs with $\ell = m =2,\,3,\,4$ and $n = 0,\,1$ as functions of $a/M$. We observe that $|C_q|$ of the fundamental modes ($n=0$) for $\ell=2,\,3,\,4$ decreases monotonically with $a/M$. We find that $|C_q|$ is generally largest for the fundamental quadrupolar mode when $a/M\lesssim0.994$, while the values of $|C_q|$ for all QNMs with $n=1$ become larger than $|C_{220}|$ close to extremality: the crossing occurs at $a/M\simeq 0.994$ for $\ell=m=2$ and $\ell=m=3$, and at $a/M\simeq 0.996$ for $\ell=m=4$.
The top panel shows that the phases of the excitation coefficients exhibit only a mild dependence on the spin up to $a/M \simeq 0.9$.
These results are consistent with those discussed previously for $\ell=m=2$, and they further underscore the importance of higher overtones for high-spin remnants.
Finally, in Fig.~\ref{fig:exc_coeff_HM} we show the real and imaginary parts of the excitation coefficients in the complex plane, with the top and bottom panels corresponding to $n=0$ and $n=1$, respectively. In both cases the trajectories spiral around the origin, and (as we observed for the $\ell=m=2$ case) the coefficients become significantly more sensitive to variations in $a/M$ for high spins ($a/M > 0.9$).

Alternative plots and some tabulated values of the excitation coefficients can be found in Appendix~\ref{app:additional_plots}.
In Appendix~\ref{app:startingpoint} we verify that the critical plunging geodesics provide a good estimate of the $C_q$'s, in the sense that the excitation coefficients depend only mildly on the starting point of the plunge.

\section{Conclusions}
In this work, we have developed the formalism to compute the excitation coefficients for particles plunging from the innermost stable circular orbit into a Kerr BH, extending previous results that considered radial plunges along the axis of symmetry~\cite{Zhang:2013ksa}.
We focused on particles plunging along equatorial critical geodesics, which are known analytically in terms of elliptic functions~\cite{Mummery:2022ana,Dyson:2023fws} and have been shown to accurately describe the transition from inspiral to plunge in the point-particle limit~\cite{Lhost:2024jmw,Faggioli:2025hff}.
Critical geodesics are only an approximation to more realistic models of the transition from inspiral to plunge for finite values of the mass ratio, but the simple numerical checks shown in Appendix~\ref{app:startingpoint} and recent work using self-force calculations~\cite{Kuchler:2025hwx} indicate that the excitation coefficients are only mildly dependent on the details of this transition.

The complex QNM amplitudes $C_q$ have an interesting dependendence on the dimensionless BH spin $a/M$.
When the BH is not close to extremality the fundamental mode with $\ell=m=2$ is dominant, but as $a/M$ increases, overtones and higher modes become more significant, and our results suggest that they may play a dominant role for near-extremal Kerr BHs.
More work is needed to characterize the relative excitation of overtones and higher modes, their detectability, and their possible role in driving gravitational turbulence~\cite{Yang:2014tla,Redondo-Yuste:2023seq,Ma:2025rnv}.

Equatorial critical plunge geodesic are particularly interesting because comparable mass binaries circularize during the inspiral. However, an extension to ``homoclinic'' and inclined orbits will provide additional insight into the excitation of QNMs in intermediate and extreme mass ratio binaries, for which the circularization mechanism is not as efficient as in the comparable mass case.  

Beyond the astrophysical implications of our results, our formalism can be combined with alternative methods to improve our theoretical understanding of the ringdown and waveform modeling. The point-particle is a good benchmark to calibrate numerical simulations and effective-one-body models, and therefore to improve current waveform templates used in gravitational-wave searches and in parameter estimation.
A comparison across different schemes for the transition from inspiral to plunge would be interesting to clarify the role of the onset of the ringdown, and whether critical plunge geodesics really are ``universal'' approximants of ringdown dynamics.

\acknowledgments
We thank G.~Carullo, M.H.-Y.~Cheung, C.~Dyson, G.~Faggioli, D.~Pere\~niguez, D.~Rossi, L.~Sberna and S.~Yi for useful discussions.
We acknowledge financial support from the EU Horizon2020 Research and Innovation Programme under the Marie Sklodowska-Curie Grant Agreement No.~101007855.
E.B. and A.M.~acknowledge support from the ITA-USA Science and Technology Cooperation program, supported by the Ministry of Foreign Affairs of Italy (MAECI). 
A.M.~acknowledges financial support from MUR PRIN Grants No.~2022-Z9X4XS and No.~2020KB33TP.
E.B.~is supported by NSF Grants No.~AST-2307146, No.~PHY-2513337, No.~PHY-090003, and No.~PHY-20043, by NASA Grant No.~21-ATP21-0010, by the John Templeton Foundation Grant No.~62840, by the Simons Foundation [MPS-SIP-00001698, E.B.], and by the Simons Foundation International [SFI-MPS-BH-00012593-02].
This work was carried out at the Advanced Research Computing at Hopkins (ARCH) core facility (\url{https://www.arch.jhu.edu/}), which is supported by the NSF Grant No.~OAC-1920103.

\appendix

\section{The Leaver method}\label{sec:Leaver_qnms}

One of the most accurate strategies to compute Kerr QNM frequencies and eigenfunctions is Leaver's method, in which the angular and radial components of the Teukolsky equation are solved in terms of continued fractions~\cite{Leaver:1985ax,Nakamura:1987zz}. 
The solution of the angular component, i.e., the spheroidal harmonics of spin $s$, can be written as 
\begin{equation}
    _{-s}S^{a\omega}_{\ell m}(z)=e^{a\omega z}(1+z)^\alpha (1-z)^\beta\sum_{n=0}^\infty a_n (1+z)^n \ ,
\label{eq:S_Leaver}
\end{equation}
where $\alpha=|m-s|/2$ and $\beta=|m+s|/2$ (the index $n$ in this expansion should not be confused with the overtone number of the QNMs). By replacing the former expansion in Eq.~\eqref{eq:S_teukolsky} for $s=-2$, it can be shown that the $a_{n}$'s satisfy a three-term recurrence relation
\begin{align}
&a_1\alpha^\theta_0+\beta^\theta_0a_0=0 \ ,\label{eq:a1_RW}\\
   &a_{n+1} \alpha^\theta_n+a_n\beta^\theta_{n}+a_{n-1}\gamma^\theta_{n}=0 \ , \quad n>0 \ ,
\label{eq:three_term_RW}\end{align}
where
\begin{align}
\alpha^\theta_n =& -2(1 + n)(1 + 2\alpha + n) \ ,\\
\beta^\theta_n =&(n-1) n+2 n (-2 a \omega +\alpha +\beta +1) \nn \\ 
&-2 a \omega  (2 \alpha +s+1)+(\alpha +\beta ) (\alpha +\beta +1)\nn \\
&-\mathcal{A}_{\ell m }-s (s+1)-a^2 \omega ^2\ , \\
\gamma^\theta_n=& 2 a \omega  (\alpha +\beta +n+s)\ .\label{eq:Leavereqs}
\end{align}
Note that $\lambda$ in Eq.~$(6.2)$ and Eq.~$(6.6)$ 
in part III of~\cite{Nakamura:1987zz}  should be replaced with $\mathcal{A}_{\ell m }$.

The solution of the radial component of the Teukolsky equation can be found in a similar way. Since we are interested in QNMs, we look for a solution $R$ with ingoing boundary conditions at the horizon $(r=r_+,\  r_\star=-\infty)$ and outgoing boundary conditions at infinity $(r= r_\star=+\infty)$. The asymptotic expansions of Eq.~\eqref{eq:R_teukolsky} give then
\begin{equation}
     R^{\rm L}_{\ell m \omega}\to
     \begin{cases}
             r^{-1-2s+i\omega} e^{i\omega r}\qquad\  r_\star \to +\infty \\
            (r-r_+)^{-s-i\sigma_+} \quad r_\star\to-\infty 
        \end{cases}\ ,
\end{equation}
where 
\begin{equation}
    \sigma_+=\frac{1}{b}\left(\omega r_+-\frac{am}{2M}\right)\ \ , \ b=2M\sqrt{1- (a/M)^2} \ .
\end{equation}
A solution of the radial Teukolsky equation with these boundary conditions is~\cite{Leaver:1985ax, Nakamura:1987zz},
\begin{equation}
R^{\rm L}_{\ell m\omega}=e^{i\omega r}(r-r_-)^{\kappa_-}(r-r_+)^{\kappa_+}\sum_{n=0}^\infty d_n \left( \frac{r-r_+}{r-r_-}\right)^n
    \label{eq:R_Leaver} \ ,
\end{equation}
where 
\begin{align}
    \kappa_-=&-1-s+2iM\omega+2iM\sigma_+ \ ,\\
     \kappa_+=&-s-2iM\sigma_+ \ , \\ 
    \sigma_+=&\left(\omega r_+-\frac{am}{2M}\right)\Big/b \ ,  b=2M\sqrt{1- (a/M)^2} \ , \\
    r_+=&M+b/2\quad\ ,\quad   r_-=M-b/2 \ .
\end{align}
The coefficients $d_n$ of the expansion are defined by a three-term recurrence relation 
\begin{align}
&d_1\alpha^r_0+\beta^r_0d_0=0 \ ,\label{eq:d1_RW}\\
   &d_{n+1} \alpha^r_n+d_n\beta^r_{n}+d_{n-1}\gamma^r_{n}=0 \ , \quad n>0 \ ,
\label{eq:three_term_RW_2}
\end{align}
where
\begin{align}
    \alpha^r_n =&(\tilde c_0+1) n+\tilde c_0+n^2 \ ,\\
    \beta^r_n=&(\tilde c_1+2) n+\tilde c_3-2 n^2 \ ,\\
    \gamma^r_n=& (\tilde c_2-3) n-\tilde c_2+\tilde c_4+n^2+2 \ , 
\end{align}
with
\begin{align}
    \tilde c_0=& -\frac{2 i \left(2\omega M^2-a m\right)}{b}-s-2 i \omega M+1\ , \\ 
    \tilde c_1=& \frac{4 i \left(2\omega M^2-a m\right)}{b}+2 i \left(b+4M\right) \omega -4\ , \\
    \tilde c_2=& -\frac{2 i \left(2\omega M^2-a m\right)}{b}+s-6 i \omega M +3 \ , \\ 
    \tilde c_3=& \frac{2(4 M\omega+ i) \left(2 \omega M^2-a m\right)}{b}+ i \left(b+4M\right) \omega \\
    & \left(-a^2+4 bM+16M^2\right) \omega ^2-2 a m\omega -\mathcal{A}_{\ell m}-s-1\ , \\
    \tilde c_4=&-\frac{(8 \omega M +2 i) \left(2\omega M^2-a m\right)}{b}-2i (2 s+3) \omega M \nn \\
    &+s-8 \omega ^2M^2+1 \ .
\end{align}
The series for $ _{-s}S^{a\omega}_{\ell m}(z)$ and $R^{\rm L}_{\ell m \omega}$, in Eqs.~\eqref{eq:S_Leaver} and ~\eqref{eq:R_Leaver}, respectively, are convergent only for a discrete set of frequencies $\omega$, for which the expansion in Eq.~\eqref{eq:R_Leaver} is a solution of the Teukolsky equation, and the following identities hold:
\begin{equation}
   \beta^i_n-\frac{\alpha^i_{n-1}\gamma^i_{n}}{\beta^i_{n-1}-} \dots \frac{\alpha^i_{0}\gamma^i_{1}}{\beta^i_0} = \frac{\alpha^i_n\gamma^i_{n+1}}{\beta^i_{n+1}-}\frac{\alpha^i_{n+1}\gamma^i_{n+2}}{\beta^i_{n+2}-}\dots  \label{eq:Gautshi_condition}
\end{equation}
with $i=(\theta, r)$. Equations \eqref{eq:Gautshi_condition} can be solved simultaneously for the QNM frequency $\omega_{\ell m n}$ and the eigenvalue $\mathcal{A}_{\ell m }$. 

For a given QNM $\omega=\omega_q$ (and $\mathcal{A}_{\ell m}=\mathcal{A}_{q}$), we can determine the coefficients $d_n$ and $a_n$ up to two normalization constants $(d_0,a_0)$.
We choose $a_0$ such that
\begin{equation}
    \int_0^\pi \sin\theta\dd \theta \ |{_{s}}S^{a\omega}_{\ell m}(\theta)|^2=1\ ,
    \label{eq:S_norm_cond}
\end{equation}
and the sign of ${_s}S^{a\omega}_{\ell m}$ is such that ${_s}Y_{\ell m}={_s}S^{a\omega=0}_{\ell m}$. We can easily determine the sign of $a_0$, noting that $_{s}S_{\ell m}^{a\omega}\to a_0 2^\beta e^{-a\omega}$ at $\theta=\pi$, and ${_s}Y_{\ell m}(\theta\to\pi^-)$ is negative if one of the following conditions is satisfied: 
\begin{itemize}
    \item $\ell$ odd and $m+s$ even,
    \item $\ell$ odd, $m+s$ odd and $m\ge s$,
    \item $\ell$ even, $m+s$ odd and $m\le s$,
\end{itemize}
while it is positive otherwise.
For the radial function, we choose $d_0=1$.

Finally, for a given QNM $\omega=\omega_q$, we construct the SN eigenfunction $X_q^{\rm L}$, which is the SN transform of $R^{\rm L}_q$. In order to obtain the function $X_q^{r_+}$ appearing in  Eq.~\eqref{eq:Iq}, $X_q^L$ must be multiplied by an appropriate constant to match the asymptotic behavior of Eq.~\eqref{eq:rescaled_SN_inf}, i.e.
\begin{align}
X_q^{r_+}&=\varpi X^{\rm L}_q  \label{eq:X_Leaver_to_X_rplus} \ ,\\
 \varpi&=-M^2 \sqrt{\frac{M}{r_+}}2^{-\frac{i a m}{r_+}+i \varepsilon -\frac{3}{2}} e^{i
   \left(\frac{a m}{2M}-2 r_+ \omega \right)}\nonumber\\
   &\times\frac{ \left(\frac{b}{M}\right){}^{\frac{i a
   m}{r_+}-4 i M \omega +1}}{\left(2 a m-ib-2\varepsilon r_+\right)\left(a m-i b - 
   \varepsilon  r_+\right)}\ .\nonumber
\end{align}
%

\section{Teukolsky and Sasaki-Nakamura amplitudes and excitation factors via the Mano-Suzuki-Takasugi method}\label{app:Teukolsky_amplitudes}

The excitation factors in the Kerr background can be computed analytically using the Mano, Suzuki and Takasugi (MST) method~\cite{Mano:1996vt}, which can be used to determine the amplitude of the Teukolsky radial solution in closed analytical form at the horizon and at infinity. For a detailed discussion of the MST method, see~\cite{Sasaki:2003xr}.

\subsection{Representation of the Teukolsky function in terms of hypergeometric functions}
\label{sec:RenAngMom}

A solution of the homogeneous Teukolsky equation with purely ingoing boundary conditions at the horizon can be written as
\begin{equation}
    R^\nu_{\rm in}=e^{i\omega b x}(-x)^{-s-i(\varepsilon+\tau)/2}
(1-x)^{i(\varepsilon-\tau)/2}p_{\rm in}^{\nu}(x) \ ,\label{eq:R_hypergeometric}
\end{equation}
where hereafter we omit the dependence of the Teukolsky and SN functions on $\ell,m$ and $\omega$, $x=(r_+-r)/b$, $\varepsilon=2 M \omega$, and $\tau=(4 \omega M^2-2ma)/b$. 
The function $p^\nu_{in}(x)$ can be expanded as
\begin{equation}
    p^\nu_{\rm in}(x)=\sum_{n=-\infty}^\infty a_{{\rm in},\,n}^\nu \ p_{n+\nu}(x)\,,
    \label{eq:seriesp}
\end{equation}
where the parameter $\nu$ is called the {\it renormalized angular momentum},
\begin{align}
 p_{\nu+n}(x)=F(n+\nu&+1-i\tau,-n-\nu-i\tau;\\
    &1-s-i \varepsilon-i\tau;x)\ ,\nonumber
\end{align}
and $F(a,b;c;z)$ is the ordinary hypergeometric function~\cite{10.5555/1098650}. Substituting Eq.~\eqref{eq:R_hypergeometric} in Eq.~\eqref{eq:R_teukolsky}, we get a three-term recurrence relation:
\begin{equation}
    \alpha_{n}^{\nu} a_{{\rm in},\,n+1}^{\nu}+\beta_n^{\nu} a_{{\rm in},\,n}^{\nu}+\gamma_n^{\nu} a_{{\rm in},\,n-1}^{\nu}=0 \ ,\label{eq:nu_3term}
\end{equation}
where
\beq \label{alpha}
\alpha_n^\nu&=&\frac{i \omega b(n+\nu+1+s+i \varepsilon)(n+\nu+1+s-i \varepsilon)}{(n+\nu+1)(2n+2\nu+3)(n+\nu+1+i\tau)^{-1}}\,,\nn\\ 
\label{beta}
\beta_n^\nu&=&-\lambda-s(s+1)+(n+\nu)(n+\nu+1)+ \varepsilon^2\,\notag\\
&+& \omega( 4\omega M^2-2ma)
+\frac{ \omega( 4\omega M^2-2ma)(s^2+ \varepsilon^2)}{(n+\nu)(n+\nu+1)}\,,\nn\\
\label{gamma}
\gamma_n^\nu&=&-\frac{i \omega b(n+\nu-s+i \varepsilon)(n+\nu-s-i \varepsilon)}{(n+\nu)(2n+2\nu-1)(n+\nu-i\tau)^{-1}}\,.
\eeq
The series~\eqref{eq:seriesp} converges for the values of $\nu$ that satisfy a continuous fraction condition similar to Eq.~\eqref{eq:Gautshi_condition}.
The numerical calculation of $\nu$ is particularly tricky, since the root of Eq.~\eqref{eq:nu_3term} is often close to a branch cut. 
This implies that carefully chosen starting values $\nu_0$ must be used to initialize root-finding algorithms based, e.g., on the standard Newton-Raphson method. 
To determine these starting values we use the ``monodromy method''~\cite{Nasipak:2024icb}, exploiting built-in functions of the Black Hole Perturbation Toolkit~\cite{BHPToolkit}. Note also that the solution is not unique: for a given value of $\nu$, $\nu\pm1$ and $-\nu$ are also roots of the recurrence relation.
Given a value of $\nu$, we can build the convergent sequence $a_{{\rm in},\, n}^\nu$, which we call equivalently $f_{n}^\nu$, and hence determine $p_{\rm in}^{\nu}(x)$ and finally $R^\nu_{\rm in}$ from Eq.~\eqref{eq:R_hypergeometric}.

\subsection{Representation of the Teukolsky function in terms of Coulomb functions}
The asymptotic amplitudes of the Teukolsky function $R_{\rm in}^\nu$, which are needed to compute the excitation factors, cannot be extracted from its representation in Eq.~\eqref{eq:R_hypergeometric}, since the series does not converge at infinity. Therefore, it is necessary to construct a different representation of the same solution of the Teukolsky equation with better convergence properties at infinity. To this aim, as a first step we note that $R_{\rm in}^\nu$ can also be written as
\be
R^\nu_{\rm in}=R_0^{\nu}+R_0^{-\nu-1} \ ,
\label{eq:R_in_RC}
\ee
where
\begin{widetext}
\begin{align}
    R_0^{\nu}=&e^{i \omega b x}(-x)^{-s-i/2(\varepsilon+\tau)}\left(1-x\right)^{(i/2)(\varepsilon+\tau)+\nu}\sum_{n=-\infty}^\infty f_n^\nu\frac{\Gamma(1-s-i \varepsilon -i\tau)\Gamma(2n+2\nu+1)}{\Gamma(n+\nu+1-i\tau)\Gamma(n+\nu+1-s-i \varepsilon )}\nonumber\\
    &\times (1-x)^nF\left(-n-\nu-i\tau,-n-\nu-s-i \varepsilon ;-2n-2\nu;\frac{1}{1-x}\right) \ .\label{eq:R0nu}
\end{align}
\end{widetext}
Remarkably, the functions $R_0^\nu$ and $R_0^{-\nu-1}$ can be written in a different form. Indeed, it is possible to represent a solution of the Teukolsky equation as
\begin{equation}
R^\nu_{{\rm C}}={\hat z}^{-1-s}\left(1-\frac{\omega b}{\hat z}\right)^{-s-i(\varepsilon+\t)/2}
f_{\n}(\hat z) \ ,\label{eq:RC_definition}
\end{equation}
where 
\begin{align}
f_{\n}(\hat z)= \sum_{n=-\infty}^{\infty}
(-i)^n\frac{(\n+1+s-i \varepsilon )_n}{(\n+1-s+i \varepsilon )_n}a^{\rm C}_n F_{n+\n}(-is-\varepsilon,z) \ ,
\label{eq:series of Rc}\end{align} 
$\hat z=\z (r- r_- )$, $(y)_{n}=\Gamma(y+n)/\Gamma(y)$, and $F_{N}(\eta,\hat z)$ is a Coulomb wave function, defined by 
\begin{align}
F_{N}(\eta,\hat z)=e^{-i\hat z}2^{N}&{\hat z}^{N+1}\frac{\G(N+1-i\eta)}{\G(2N+2)}\times\nonumber\\
&\Phi(N+1-i\eta,
2N+2;2i\hat z) \ .
\label{eq:defcoulomb}
\end{align}
Here $\Phi(\a,\b;\hat z)$ is the confluent hypergeometric function, which is regular at $\hat z=0$~\cite{10.5555/1098650}. 
Substituting Eq.~\eqref{eq:RC_definition} in the radial component of the Teukolsky equation~\eqref{eq:R_teukolsky}, we obtain a three-term recurrence relation similar to Eq.~\eqref{eq:nu_3term}, with the parameter $\nu$ being the same for both $R^\nu_{\rm C}$ and $R^\nu_{\rm in}$~\cite{mano_analytic_1996}.

The two series~\eqref{eq:R0nu} and \eqref{eq:RC_definition} are both defined for $\omega b<\hat z<\infty$ and are solutions of the Teukolsky equation. Since they have the same behavior for large (finite) $\hat z$,  they are proportional to each other:
\be
R_0^\nu=K_\n R^\n_{\rm C} \, ,
\ee
where
\begin{widetext}
\begin{eqnarray}
K_{\n}
&=&\frac{e^{i\omega b}(2\omega b)^{s-\n-N}2^{-s}i^{N}\Gamma(1-s-i \varepsilon-i\tau)\Gamma(N+2\n+2)}
{\Gamma(N+\n+1-s+i \varepsilon )\Gamma(N+\n+1+i\tau)\Gamma(N+\n+1+s+i \varepsilon)}\nonumber \\
&&\times\left(\sum_{n=N}^{\infty}(-1)^{n}\frac{\Gamma(n+N+2\n+1)}{(n-N)!}
\frac{\Gamma(n+\n+1+s+i \varepsilon)\Gamma(n+\n+1+i\tau)}
{\Gamma(n+\n+1-s-i \varepsilon )\Gamma(n+\n+1-i\tau)}f_{n}^{\n}\right)\nonumber \\
&&\times\left(\sum_{n=-\infty}^{N}\frac{(-1)^{n}}{(N-n)!(N+2\n+2)_{n}}
\frac{(\n+1+s-i \varepsilon )_{n}}{(\n+1-s+i \varepsilon )_{n}}f_{n}^{\n}\right)^{-1},
\end{eqnarray}
\end{widetext}
$N$ can be any integer, and the factor $K_{\nu}$ is independent of the choice of $N$.
While the representation~\eqref{eq:R_hypergeometric} is well behaved at the horizon, we can now find a representation of the function $R_{\rm in}^\nu$, i.e.,
\begin{eqnarray}
\label{eq:secondRin}
R^\nu_{\rm in}=K_{\n}R_{{\rm C}}^{\n}+K_{-\n-1}R_{{\rm C}}^{-\n-1} \ ,
\end{eqnarray}
which is well behaved at infinity. 
\subsection{Teukolsky amplitudes}\label{subsec:TA}
We can now determine the amplitudes $B^{{\rm trans}}$, $B^{{\rm inc}}$ and $B^{{\rm ref}}$ of $R^\nu_{\rm in}$ by studying the asymptotic behavior 
\beq  R^\nu_{\text{in}}\to \left\{
\begin{array}{l}
B^{\text{trans}}\Delta^2e^{-ip_\omega r_\star}~{\rm as}~r\to r_+\,,\\
r^3B^{\text{ref}}e^{i\omega
r_\star}+r^{-1}B^{\text{inc}}e^{-i\omega r_\star}~{\rm as}~r\to +\infty\  .\\
\end{array}
\right.
\label{eq:RinAsy} 
\eeq
Expanding Eq.~(\ref{eq:R_hypergeometric}) in the limit $r\rightarrow r_{+}$ and Eq.~(\ref{eq:secondRin}) in the limit $r\rightarrow\infty$, we find 
\begin{subequations}
\begin{align}
B^{{\rm trans}}
=&\,b^{2s}e^{i  b/(2M) \varepsilon_+\left(1+2\frac{\ln [b/(2M)]}{1+b/(2M)}\right)}
\sum_{n=-\infty}^{\infty}f_{n}^{\n} \ ,\\
B^{{\rm inc}}
=&\,\omega^{-1}\left[K_{\n}-ie^{-i\pi\n}
\frac{\sin\pi(\n-s+i \varepsilon )}{\sin\pi(\n+s-i \varepsilon )}K_{-\n-1}\right]\nn\\
&\,\times A_{+}^{\n}e^{-i\left(\varepsilon\ln\varepsilon-\frac{2M-b}{2}\omega\right)} \ , \\
B^{{\rm ref}}
=&\,\omega^{-1-2s}[K_{\n}+ie^{i\pi\n}K_{-\n-1}]\nn\\
&\times A_{-}^
{\n}e^{i\left(\varepsilon\ln\varepsilon-\frac{2M-b}{2}\omega\right)} \ ,   
\end{align}
\label{eq:asymp_amp}
\end{subequations}
where $\varepsilon_+=(\varepsilon+\tau)/2$ and
\begin{align}
A_{+}^{\n}=&\,2^{-1+s-i \varepsilon }e^{-\frac{\pi}{2}\varepsilon}e^{\frac{\pi}{2}i(\n+1-s)}
\frac{\G(\n+1-s+i \varepsilon  )}{\G(\n+1+s-i \varepsilon )}\sum_{n=-\infty}^{+\infty}f_{n}^{\n} \ ,
\nonumber \\
A_{-}^{\n}=&\,2^{-1-s+i \varepsilon }e^{-\frac{\pi}{2}\varepsilon}e^{\frac{-\pi}{2}i(\n+1+s)}\nn \\
&\times \sum_{n=-\infty}^{+\infty}(-1)^{n}\frac{(\n+1+s-i \varepsilon )_{n}}{(\n+1-s+i \varepsilon )_{n}}
f_{n}^{\n} \ .
\end{align}
It is also convenient to introduce the normalized amplitudes of the Teukolsky function, $A^{\rm T\, in}=B^{\rm inc}/B^{\rm trans}$ and $A^{\rm T\, out}=B^{\rm ref}/B^{\rm trans}$, which are independent of the overall normalization.

\subsection{Sasaki-Nakamura amplitudes}
The normalized SN amplitude $A_q^{\rm out}$ appearing in Eq.~\eqref{eq:Iq}  can be computed from the Teukolsky amplitudes as follows. 
Let $X^{\rm in}$ be the solution of the SN equation satifying ingoing boundary conditions, i.e.,
\begin{equation}
    X^{\rm in}\to \begin{cases}
        A^{\rm ref} e^{i\omega r_\star}+A^{\rm inc} e^{-i\omega r_\star} &r_\star \to +\infty \\
        A^{\rm trans} e^{-ip_\omega r_\star}  &r_\star \to -\infty
    \end{cases}\ .
    \label{eq:def_SN_in}
\end{equation}
The coefficients $A^{\rm ref}$, $A^{\rm inc}$ and $A^{\rm trans}$ are 
related to the amplitudes of the corresponding 
Teukolsky equation $B^{\rm ref}$, $B^{\rm inc}$ and $B^{\rm trans}$ in Eqs.~\eqref{eq:asymp_amp} by (see e.g.~\cite{Sasaki:2003xr}):
\begin{subequations}\label{eq:amplitudesSN}
\begin{eqnarray}
    A^{\rm inc}&=& -4 \omega^2 B^{\rm inc}\ ,\\
    A^{\rm ref}&=& -\frac{c_0}{4 \omega^2} B^{\rm ref}\ \ ,\\
    A^{\rm trans}&=&d \ B^{\rm trans}\ ,
\end{eqnarray}
\end{subequations}
where 
\begin{align}
    d&=\sqrt{ 2M r_+} \left[8 M^2-12 i a m M-4 a^2 m^2\right.
    \nn\\
    &\left.+r_+\left(12 i a m+16 a m M
   \omega +24 i M^2 \omega -16 M\right)\right.\nn \\
   &\left.+r_+^2
   \left(-16 M^2 \omega ^2-24 i M \omega +8\right)\right]\,.
\end{align}
We then define the normalized amplitudes $A^{\rm in}=A^{\rm inc}/A^{\rm trans}$ and $A^{\rm out}=A^{\rm ref}/A^{\rm trans}$. At the QNM frequencies, $\omega=\omega_q$, $A_q^{\rm out}=A^{\rm out}$ and $A_q^{\rm in}=0$.
\subsection{The Sasaki-Nakamura excitation factors}\label{sec:excitation_factors}
\begin{figure}[t]
    \centering
    \includegraphics[width=0.98\linewidth]{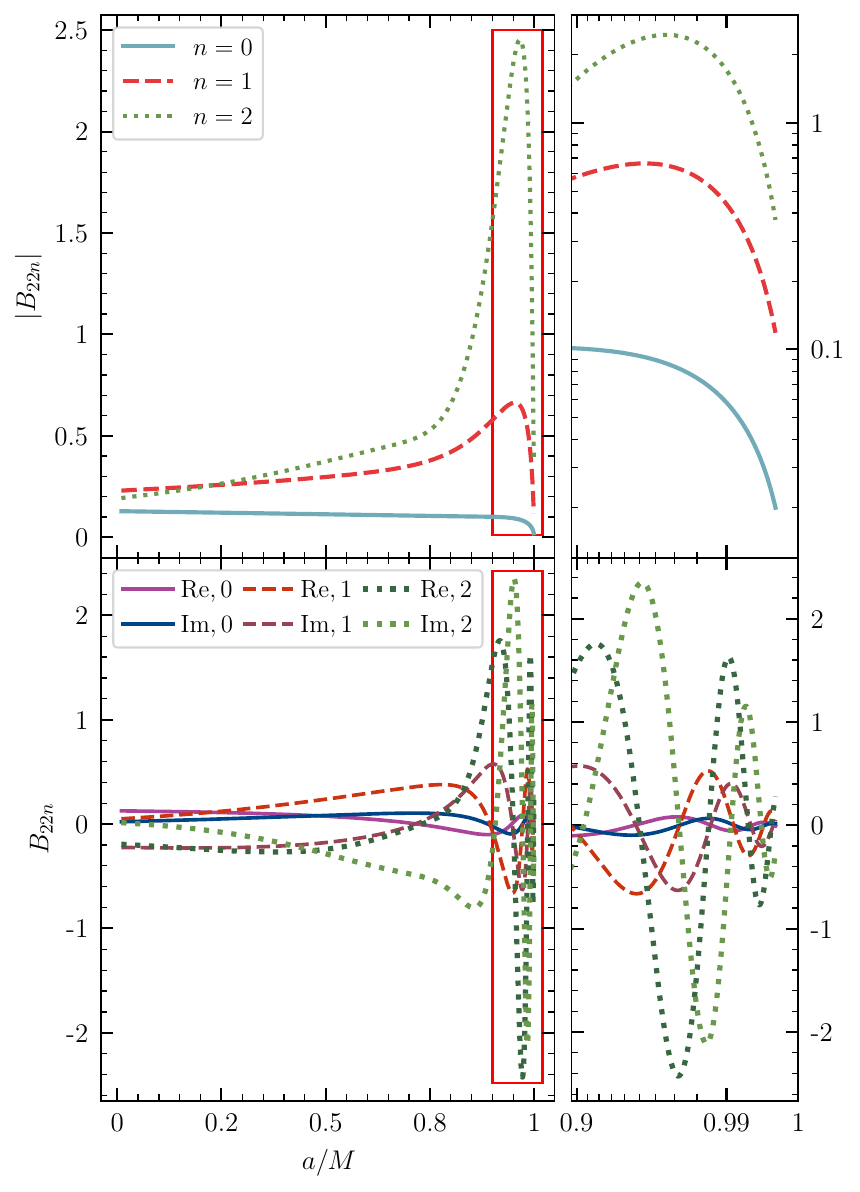}
    \caption{Excitation factors for $\ell=m=2$ and $n=0,\,1,\,2$ as functions of $a/M$. The absolute value is shown in the top panels, the real and the imaginary parts in the bottom panels.
    The right panels zoom in $a>0.9\ M$, which is marked by a rectangle in the left panels. In both panels, numbers in the legend refer to the overtone number $n$.
    }
    \label{fig:exc_fac}
\end{figure}

The excitation factors of the Teukolsky function, $B^{\rm T}_q$, can be written in terms of the normalized amplitudes defined in Sec.~\ref{subsec:TA}:
\begin{equation}
    B_q^{\rm T}=\frac{A^{\rm T\,out}}{2 i\omega_q\alpha_q^{\rm T} } \ ,\qquad \alpha_q^{\rm T}=\frac{\dd A^{\rm T\, in}}{\dd\omega }\bigg|_{\omega=\omega_q} \ .\label{eq:Teukolsky_excitation_factors}
\end{equation}
From $B^{\rm T}_{q}$, using Eqs.~\eqref{eq:amplitudesSN}, we can compute the excitation factors of the rescaled SN functions $B_q$, defined in Eq.~\eqref{eq:excitation_factors}:

\begin{equation}
    B_q=\,B_q^{\rm T}\frac{c_0}{16\omega^4} \ ,\label{eq:excitationfactorsSN}
\end{equation}
where  $c_0=\lambda(\lambda+2)-12 Mi\omega-12 a\omega(a\omega -m)$ (Eq.~\eqref{eq:c_0}). Note that $B_q$ coincide with the excitation factors of the (non-rescaled) SN function (cf. Eq.~(35) in~\cite{Zhang:2013ksa}).
In Fig.~\ref{fig:exc_fac} we show the excitation factors for $\ell=m=2$ and for the first three overtones ($n=0,\,1,\,2$). In the top panels we show the absolute value, while in the bottom panels we show the real and imaginary parts. The left panels show the whole range of the spin, while the right panels zoom in on highly rotating BHs ($a/M>0.9$), i.e., the region highlighted by a red rectangle in the left panels.
While $|B_{220}|$ is monotonically decreasing as a function of $a/M$, $|B_q|$ for the first (second) overtone increases up to $a/M\simeq 0.95$ ($a/M\simeq 0.96$), where it has a peak, and then decreases. The peak for $n=2$ is almost $4$ times larger than the peak for $n=1$. For any spin we have $|B_{220}|<|B_{221}|<|B_{222}|$.
Since $B_q$ is independent of the normalization by definition, the real and the imaginary parts are meaningful quantities. 
They do not show much variation up to $a\simeq 0.8\ M$, where they start to oscillate rapidly around $0$ (see the bottom  right panel). These oscillations have larger amplitude for larger values of $n$. This is consistent with the large peak in the absolute value of $B_{22n}$ for $n>0$. 
An inspection of the rescaled SN amplitude $A^{\rm out}_q$ shows that  the large peak in $|B_q|$ is due to a peak in $|A^{\rm out}_q|$.
Contrary to the excitation factors, the $C_q$'s do not depend explicitly on $A_q^{\rm out}$ (see Eq.~\eqref{eq:Cq_def}), therefore the absolute values of the excitation coefficients do not exhibit a peak and the real and imaginary parts do not rapidly oscillate at high spins. 

\begin{figure}[t]
    \centering
    \includegraphics[width=0.98\linewidth]{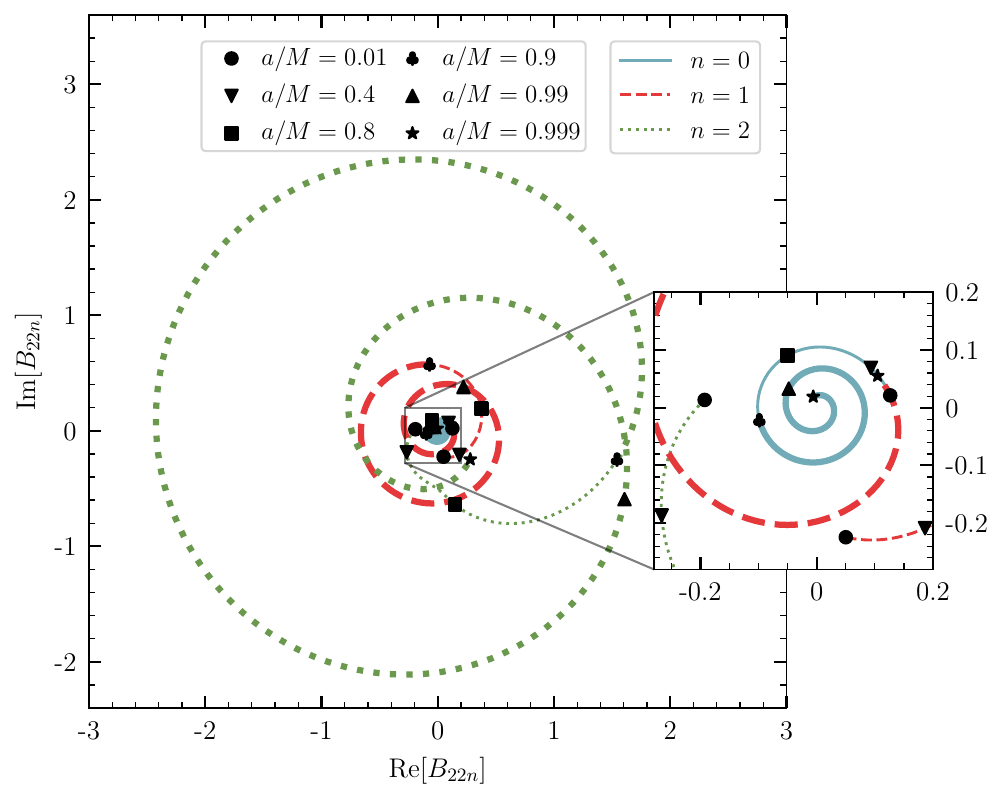}
    \caption{Same as fig.~\ref{fig:exc_coeff_re_vs_im} but for the excitation factors $|B_q|$.}
    \label{fig:exc_fac_complex_plane}
\end{figure}

\begin{figure}[t]
    \centering
    \includegraphics[width=0.98\linewidth]{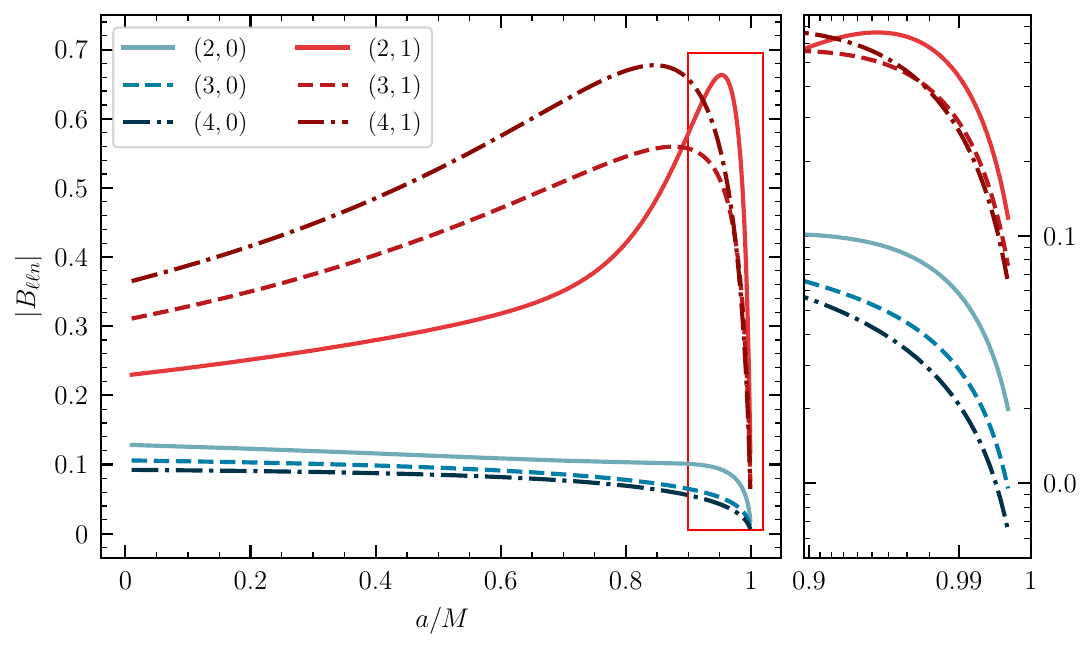}
    \caption{Excitation factors for $\ell=m=2,\,3,\,4$ and $n=0,\,1$
    as functions of $a/M$. The pairs $(\ell=m,n)$ are shown in the legend.
    }
    \label{fig:exc_fac_HM}
\end{figure}

In Fig.~\ref{fig:exc_fac_complex_plane} we show the trajectory of the $B_q$'s, parametrized by $a/M$, in the complex plane.
All of the excitation factors spiral around the origin. Thicker lines refer to $a>0.9 \ M$, and selected values of the spin are marked by the symbols shown in the legend.
The inset zooms in around the origin to better show the fundamental mode ($n=0$). The $n=2$ overtone spans a much larger region of the complex plane (consistently with the large amplitude peak shown in Fig.~\ref{fig:exc_fac}).
In Fig.~\ref{fig:exc_fac_HM} we plot $|B_{\ell \ell n}|$ for $2\le \ell\le4$ and $n=0,\,1$. 
For the fundamental mode this quantity decreases monotonically for any $\ell$, while for $n=1$ is has a maximum around $a/M \simeq0.9$. Overall, the excitation factors for the first overtone, $|B_{\ell\ell 1}|$, are more sensitive to $\ell$ than those for the fundamental mode, $|B_{\ell\ell 0}|$. While the relative ordering of $|B_{\ell\ell 1}|$ changes and $|B_{221}|$ becomes dominant at $a/M\simeq 0.9$, we observe that $|B_{220}|> |B_{330}|> |B_{440}|$ for any value of $a/M$.

\begin{figure}[t]
    \centering
    \includegraphics[width=0.81\linewidth]{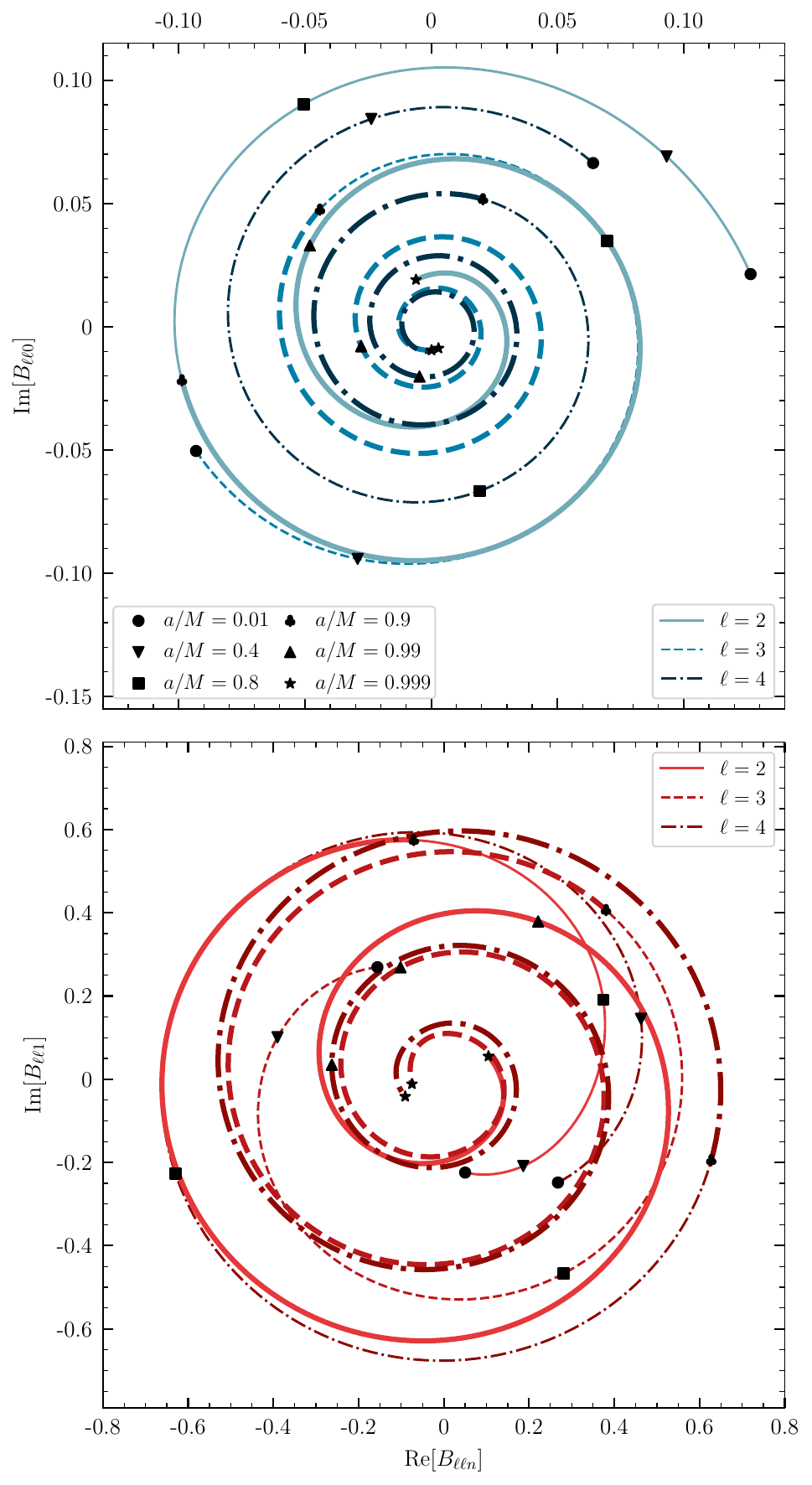}
    \caption{Same as Figure~\ref{fig:exc_coeff_HM} but for the excitation factors $|B_q|$.
    }
    \label{fig:app:exc_fac_HM_complex_plane}
\end{figure}

\begin{figure}[t]
    \centering
    \includegraphics[width=0.98\linewidth]{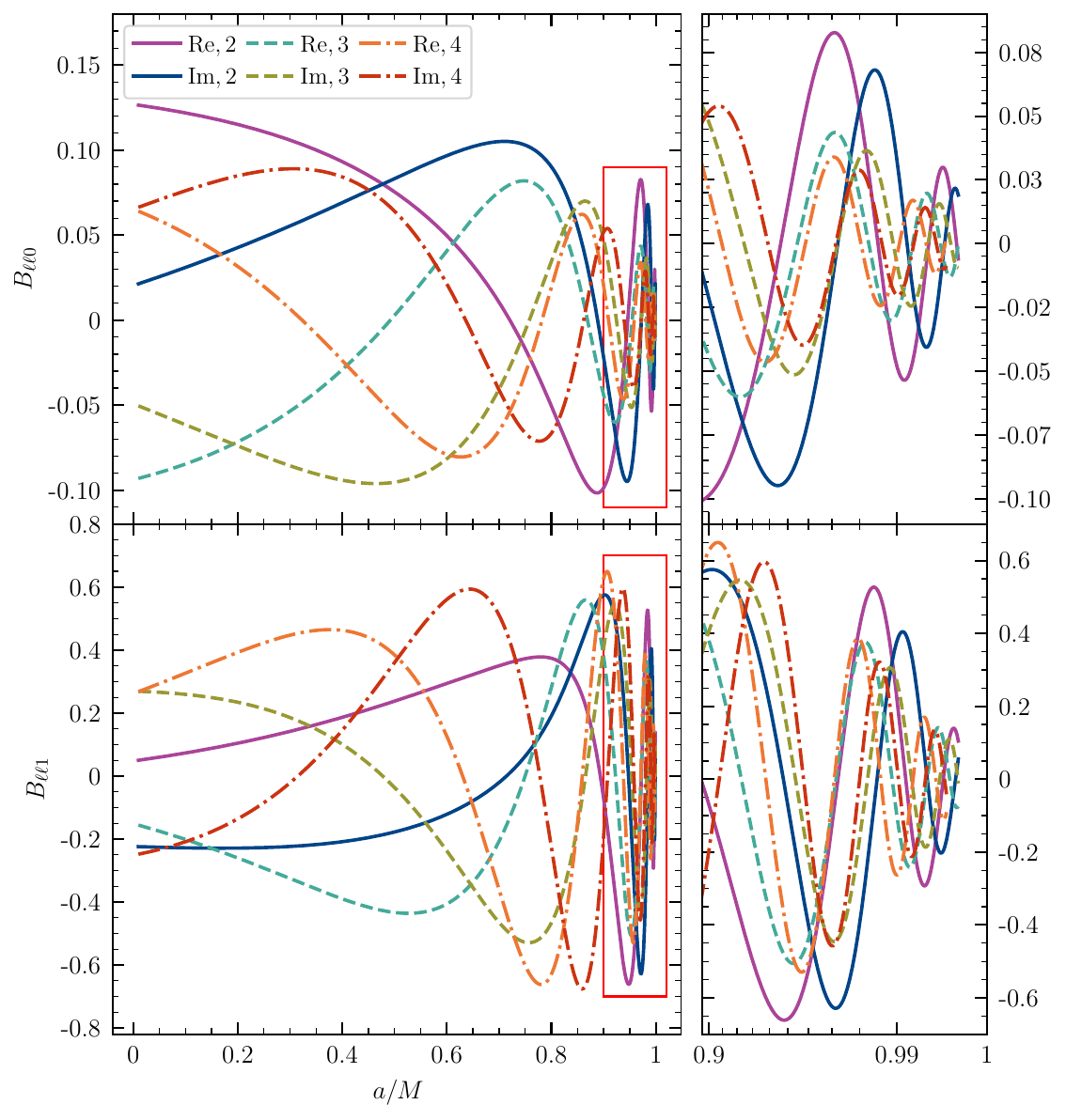}
    \caption{ Excitation factors for $\ell=m=2,\,3,\,4$ and $n=0$ (top panel) and $n=1$ (bottom panel) 
    as functions of $a/M$. Numbers in the legend refer to the harmonic index $\ell=m$.
    }
    \label{fig:app:exc_fac_HM_ReIm}
\end{figure}

The complex plane trajectories of $B_{\ell\ell n}$ for $2\le\ell\le 4$ and $n=0,\,1$ are shown in Fig.~\ref{fig:app:exc_fac_HM_complex_plane}, while the real and imaginary parts are shown as functions of $a/M$ in Fig.~\ref{fig:app:exc_fac_HM_ReIm}.
In both figures, the top panel refers to $n=0$, and the bottom panel refers to $n=1$. 
In the complex plane, the excitation factors have features similar to the excitation coefficients shown in Fig.~\ref{fig:app:exc_fac_HM_complex_plane}, and in particular they change very rapidly for near-extremal spins. 
In Fig.~\ref{fig:app:exc_fac_HM_ReIm} we see that $\Re[B_{330}]$ has only one extremum at $a/M\sim 0.8$ in the range $0\le a/M < 0.9$, while $\Re[B_{440}]$, $\Im[B_{440}]$ and $\Im[B_{330}]$ have two extrema in that range. For near-extremal BHs ($a/M> 0.9$), the real and imaginary parts oscillate rapidly for all of these modes.

%
\section{Regularization procedure for \texorpdfstring{$I_q$}{I_q}}\label{app:Iq_reg}
As discussed in Sec.~\ref{sec:excitationcoefficients}, the integral $I_q$ defined in Eq.~\eqref{eq:Iq} diverges. This is due to the behavior of the functions $X_q$ and $\xi_q$ at $r\to r_+$.
Indeed, 
\begin{align}
    X^{r_{+}}&= (r-r_+)^{-2iM\sigma_+}\sum_{n=0}^\infty x_n(r-r_+)^n  \; ,\\
    e^{-i\xi}&= (r-r_+)^{-2iM\sigma_+} \sum_{n=0}^\infty\xi_n (r-r_+)^n \; ,
\end{align}
while the function $W(r)$ is regular at the horizon (see Appendix~\ref{app:W_generic}):
\begin{equation}
    W(r)=\sum_{p=0}^\infty w_p  (r-r_+)^p\; .
\end{equation}
Thus the integrand in Eq.~\eqref{eq:Iq} behaves as
\begin{equation}
(r-r_+)^{-4iM\sigma_+}\sum_{j=0}^\infty(r-r_+)^{j} \iota_j\,, 
    \label{eq:iq_rp}
\end{equation}
where the coefficients $\iota_j$ are determined by the expansion of the integrand near $r_+$. We have 
\begin{equation}
    \sigma_+=\frac{\omega r_+- am/(2M)}{r_+-r_-}\; ,
\end{equation}
and the term $(r-r_+)^{4M\Im[\omega]r_+/(r_+-r_-)}$ in Eq.~\eqref{eq:iq_rp} leads to a non-integrable divergence for $r\to r_+$ and $\Im[\omega]< -(r_+-r_-)/4Mr_+$.
The physical meaning of this divergence and the techniques to get rid of it have been extensively discussed in the literature~\cite{Detweiler:1979xr,Leaver:1986gd,Sun:1988tz}. The most recent results indicate that this is related to an improper choice of the Green's function that does not enforce causality~\cite{DeAmicis:2025xuh}. Here we follow a standard procedure and we regularize the integral $I_q$  by adding a divergent surface term, i.e., an integral of the form
\begin{equation}    
\int_{r_+}^{\infty}\dd r \ \mathscr{B}(r) \ ,
\end{equation}
where $\mathscr{B}$ is a total derivative, that diverges on the horizon like the integrand of $I_q$ and vanishes at infinity.
As discussed in~\cite{DeAmicis:2025xuh} and~\cite{Sun:1988tz}, this procedure is equivalent to modifying the integration contour in the spatial variable.
In practice,
we define the function $\mathscr{B}(r)$ as
\begin{equation}
    \mathscr{B}=\frac{\dd }{\dd r}\sum_{j=0}^N\left(\frac{\bar b_j}{\zeta_q+j+1} (r-r_+)^{\zeta_q+j+1}e^{-(r-r_+)}\right)\,,
    \label{eq:bounds_funcs}
\end{equation}
which removes the divergent part of the original integral in Eq.~\eqref{eq:Iq} for a proper choice of the coefficients $\bar b_j$, $\zeta_q$ and of the integer $N$.
We choose
\begin{equation}
    \zeta_q=-4 M i \sigma_+\,,
\end{equation}
and choose $N$ as the smallest integer larger than $4\Im[\omega] r_+/(r_+-r_-)$. Remarkably, adding more terms to the sum would not affect the result, since the terms with $j>N$ give vanishing contribution to the integral.
The actual integral to be computed in Eq.~\eqref{eq:Iq} is thus the regularized quantity
\begin{align}
&\mathcal{I}_q
=\frac{\sqrt{c_0}}{A_{\rm out}}\nonumber\\
&\int_{r_+}^\infty \dd r'\  \left(X^{r_{+}}(r')\frac{W(r')}{r'^2(r'^2+a^2)^{1/2}}e^{- i\xi(r')} +\mathscr{B}(r)\right)\;
.\label{eq:Iq_regularized}
\end{align}
For a particle plunging from a finite radius $r_p$, $W(r)$ has compact support:
\begin{equation}
\begin{cases}
    W(r)\ne 0 & r_+\le r\le r_p \, ,\\ 
    W(r)=0 & r> r_p \, ,
\end{cases}
\end{equation}
and the integral $\mathcal{I}_q$ can be recast as
\begin{align}
\mathcal{I}_q=
&\frac{\sqrt{c_0}}{A_{\rm out}}\nonumber\\
&\int_{r_+}^{r_p} \dd r'\  \left(X^{r_{+}}(r')\frac{W(r')}{r'^2(r'^2+a^2)^{1/2}}e^{- i\xi(r')} +\mathscr{B}(r)\right)\nonumber\\
&-B(r_p)\; ,\label{eq:Iq_regularized_compact}
\end{align}
where $B(r)$ is the primitive function of $\mathscr{B}(r)$.
In the specific case of a particle plunging from the ISCO, the orbit starts at $r_p=r_{\rm I}$.
\section{Sasaki-Nakamura source term for generic trajectories} \label{app:W_generic}
The SN source term $\mathcal{S}_{\ell m\omega}$ can be determined in terms of the function $W(r)$ defined in Eq.~\eqref{eq:W_def}, which is related to the source term $\lmo{T}$ of the Teukolsky equation via Eq.~\eqref{eq:W_T_link}. Here we determine $W(r)$ for a general geodesic orbit. A similar calculation (with a few typos) can be found in Ref.~\cite{Watarai:2024huy}.

We can split  $W$ in three parts,
\begin{equation}
    W = W_{nn} + W_{\overline{mm}} + W_{\overline{m}n}\:,
\end{equation}
in analogy with the splitting of the Teukolsky source term in Eq.~\eqref{eq:T_lmw}.
From Eq.~\eqref{eq:W_T_link}, using Eqs.~\eqref{eq:T_fin} and $\eqref{eq:C_T_tetrad}$, we find that $W_{nn}, W_{\overline{mm}},$ and $W_{\overline{m}n}$ satisfy the following equations:
\begin{widetext}
    \begin{subequations}
\begin{align}
     \frac{\sqrt{2\pi}}{m_p}\derr{W_{nn}}{r} &= \frac{r^2}{2\rho\Delta^2}\left| \derivative{r}{\tau} \right| \left( 1 - \frac{P}{\sqrt{\mathcal{R}}} \right)^2 L^\dagger_1\left[ \rho^{-4} L^\dagger_2 (\rho^3 S) \right] \mathrm{e}^{i\zeta}\:, \label{eq:diff_W_nn} \\
    \begin{split}
        \frac{\sqrt{2\pi}}{m_p}\derr{W_{\overline{m}n}}{r} &= \int_{\mathbb{R}} \dd{t} \:\mathrm{e}^{i\zeta}\left\{ -\frac{r^2 \bar{\rho}}{\rho^2} L^\dagger_2(\rho\bar{\rho}S)J_+\left[ \rho \frac{\dd{r}}{\dd{t}} \frac{\Sigma}{\Delta} \left( 1 - \frac{P}{\sqrt{\mathcal{R}}}  \right)w^{(1)}_{\overline{m}n} \delta(r-r(t)) \right] \right\}\\
        &- \mathrm{sgn}\left( \derivative{r}{\tau} \right) \frac{r^2 \rho}{2}L^\dagger_2 \left[ \rho^3 S (\bar{\rho}^2\rho^{-4})' \right] \frac{\Sigma}{\Delta} \left( 1-\frac{P}{\sqrt{\mathcal{R}}} \right) w^{(1)}_{\overline{m}n} \mathrm{e}^{i\zeta}\:,\label{eq:diff_W_mn}
    \end{split} \\
    \frac{\sqrt{2\pi}}{m_p} \derr{W_{\overline{mm}}}{r} &= \int_{\mathbb{R}} \dd{t} \:\mathrm{e}^{i\zeta} S 
    \left\{ r^2\rho^3 J_+ 
    \left[ \rho^{-4} J_+ 
    \left( \frac{\rho\bar{\rho}^2}{2} \left( \derivative{t}{\tau} \right)^{-1} \delta(r-r(t)) \left(w^{(1)}_{\overline{m}n}\right)^2 \right) \right] 
    \right\}\:, \label{eq:diff_W_mm}
\end{align} 
\label{eq:W_Watarai_def}
\end{subequations}
\end{widetext}

where $\zeta=\zeta(r)$ and the quantities $w_{\overline{m}n}^{(1)}$ are defined by
\begin{align}
    \zeta(r)&= \int \frac{K}{\Delta} \, \dd r-m \phi+\omega  t\,,\\
    w^{(1)}_{\overline{m}n} &= -\left[ \pm\sqrt{\Theta} + i\sin\theta\left( a\EN-\frac{\ANG}{\sin^2\theta} \right) \right]\:,
\end{align}
where the functions $P$ and $\mathcal{R}$ are defined in Sec.~\ref{sec:sourceterm} and Sec.~\ref{sec:geodesics}, respectively. The function $\Theta(\theta)=Z(z)/\sqrt{1-z^2}$ (see Eq.~\eqref{eq:geo:polareom}) determines the polar motion, and it vanishes for an equatorial infall.
Integrating Eqs.~\eqref{eq:W_Watarai_def} by parts, we get
\begin{subequations}
    \begin{align}
    &\frac{\sqrt{2\pi}}{m_p}W_{nn}(r)=\:f_0(r)\:\mathrm{e}^{i\zeta(r)} + \int^\infty_{r} f_1(r_1)\:\mathrm{e}^{i\zeta(r_1)}\dd{r}_1 + \nn \\
    &\int^\infty_{r} \dd{r}_1 \int^\infty_{r_1} f_2(r_2)\:\mathrm{e}^{i\zeta(r_2)}\dd{r}_2\:, \label{eq:W_nn}\\
    &\frac{\sqrt{2\pi}}{m_p}W_{\overline{m}n}(r) =\:g_0(r)\:\mathrm{e}^{i\zeta(r)} + \int^\infty_{r} g_1(r_1)\:\mathrm{e}^{i\zeta(r_1)}\dd{r}_1 \nn\\
    &+ \int^\infty_{r} \dd{r}_1 \int^\infty_{r_1} g_2(r_2)\:\mathrm{e}^{i\zeta(r_2)}\dd{r}_2\:, \label{eq:W_mn} \\
    &\frac{\sqrt{2\pi}}{m_p}W_{\overline{mm}}(r) =\:h_0(r)\:\mathrm{e}^{i\zeta(r)} + \int^\infty_{r} h_1(r_1)\:\mathrm{e}^{i\zeta(r_1)}\dd{r}_1 \nn \\
    &+\int^\infty_{r} \dd{r}_1 \int^\infty_{r_1} h_2(r_2)\:\mathrm{e}^{i\zeta(r_2)}\dd{r}_2\: \label{eq:W_mm},
\end{align}
\label{eq:W_}
\end{subequations}
where 
    \begin{subequations}
    \begin{align}
    \label{eq:f_0}
    f_0 &= -\frac{1}{\omega^2}w_{nn}\:,\\
    f_1 &= -\frac{1}{\omega^2} \left[ w'_{nn} + i\eta w_{nn}  \right]\:,\\
    f_2 &= \frac{i}{\omega} \left[\left( w'_{nn} + i\eta w_{nn} \right) H+ w_{nn} H'  \right]\:,
    \end{align}
    \end{subequations}
    \begin{subequations}
    \begin{align}
    g_0 &= \frac{i}{\omega} w^{(1)}_{\overline{m}n} \frac{w^{(2)}_{\overline{m}n}}{\bar{\rho}(r^2+a^2)} \mathrm{sgn}\left( \derivative{r}{\tau} \right)\:,\\
    g_1 &= \frac{i}{\omega} \mathrm{sgn}\left( \derivative{r}{\tau} \right)w^{(1)}_{\overline{m}n} \left[ -w^{(3)}_{\overline{m}n} + \left( \frac{w^{(2)}_{\overline{m}n}}{\bar{\rho}(r^2+a^2)} \right)' \nn \right.\nonumber\\
    &\left.+ \frac{w^{(2)'}_{\overline{m}n}}{\bar{\rho}(r^2+a^2)} + i\eta \frac{w^{(2)}_{\overline{m}n}}{\bar{\rho}(r^2+a^2)} \right] \:,\\
    g_2 &= -\frac{i}{\omega} w^{(1)}_{\overline{m}n} \left[ \left( w^{(3)}_{\overline{m}n} - \frac{w^{(2)'}_{\overline{m}n}}{\bar{\rho}(r^2+a^2)} \right)' \right.\nn \nonumber\\
    &\left. + i\eta \left( w^{(3)}_{\overline{m}n} - \frac{w^{(2)'}_{\overline{m}n}}{\bar{\rho}(r^2+a^2)} \right)  \right] \mathrm{sgn}\left( \derivative{r}{\tau} \right)\:,
    \end{align}
    \end{subequations}
    \begin{subequations}
    \begin{align}
    h_0 &= \frac{r^2\bar{\rho}^2}{2}\left| \derivative{r}{\tau} \right|^{-1} S \left(w^{(1)}_{\overline{m}n}\right)^2\:,\\
   \label{eq:h_} h_1 &= \frac{\rho\bar{\rho}^2}{2} \left| \derivative{r}{\tau} \right|^{-1} S \left(w^{(1)}_{\overline{m}n}\right)^2 \left[ \left(\frac{r^2}{\rho}\right)' + \rho^{-4} \left( \rho^3 r^2 \right)'
 \right]\:,\\
    h_2 &= \frac{\rho\bar{\rho}^2}{2} \left| \derivative{r}{\tau} \right|^{-1} S \left(w^{(1)}_{\overline{m}n}\right)^2 \left[ \rho^{-4}\left( \rho^3 r^2 \right)' \right]'\:,
\end{align}
\end{subequations}

with
\begin{subequations}
    \begin{align}
    \label{eq_w_nn}
    w_{nn} &= \frac{r^2}{2 \rho (r^2+a^2)^2} \left| \derivative{r}{\tau} \right| L_1^{\dagger} \left[ \rho^{-4} L_2^{\dagger} (\rho^3 S) \right]\:,\\
    w^{(2)}_{\overline{m}n} &= \frac{r^2 \bar{\rho}}{\rho^2} L^\dagger_2 \left[ \rho\bar{\rho} S \right]\:,\\
    w^{(3)}_{\overline{m}n} &= \frac{r^2}{2\bar{\rho} (r^2+a^2)} L^\dagger_2 \left[ \rho^3 S \left( \bar{\rho}^2 \rho^{-4} \right)' \right]\:,\\
    H&= v' - \frac{a(a\EN\sin^2\theta - \ANG)}{\sqrt{\mathcal{R}}}\nn \\
    &=\frac{r^2+a^2}{\Delta}\left(1+\mathrm{sign}\left(\frac{\dd r}{\dd \tau}\right)\frac{ P }{\sqrt{\mathcal{R}}}\right)
    \end{align}
\end{subequations}

and 
\begin{align}
v &= t + r_\star\:,\\
\tilde{\phi} &= \phi + \int^{r} \frac{a}{\Delta} \dd{r}\:.
\end{align}
To compute $f_i, g_i$, and $h_i$ $(i=0,\,1,\,2)$ we use the following properties of an arbitrary function $f(r)$:
\begin{align}
    \label{eq:identity_1}
    f H\mathrm{e}^{i\zeta} &= -\frac{i}{\omega} \left[\left( f\mathrm{e}^{i\zeta} \right)' - \left(f' + i\eta f \right) \mathrm{e}^{i\zeta} \right]\:, \\
    \label{eq:identity_2}
     J_+\left[ f\right]  &=  \left[ f\ \mathrm{exp}\left(i\int^r \frac{K}{\Delta}\dd{r} \right)\right]'\mathrm{exp}\left(-i\int^r \frac{K}{\Delta}\dd{r} \right)\,,
\end{align}
where
\begin{equation}
\begin{split}
    \eta(r) &= \frac{a\omega (a\hat{E}\sin^2\theta - \hat{L})}{\sqrt{\mathcal{R}}} - m\tilde{\phi}'\\ &= \left( a\omega - \frac{m}{\sin^2\theta} \right)\frac{a\hat{E}\sin^2{\theta}-\hat{L}}{\sqrt{\mathcal{R}}} - \frac{am}{\Delta} \left( 1 - \frac{P}{\sqrt{\mathcal{R}}} \right)\:.
\end{split}
\end{equation}
For equatorial orbits, the angular terms involving $S$ become
\begin{subequations}
    \begin{align}
    L_1^{\dagger} &\left[ \rho^{-4} L_2^{\dagger} (\rho^3 S) \right]\Big|_{\theta=\pi/2} =\:2r \bigg\{\left(-m+a\omega-\frac{ia}{r}\right)\big[ S_1 \nonumber\\
    &\qquad \qquad \qquad \qquad + (-m +a\omega)S_0 \big] - \frac{\lambda}{2}S_0\bigg\}\,, \\
    L_2^{\dagger} &\left[ \rho \bar{\rho} S \right]\Big|_{\theta=\pi/2} = \frac{1}{r^2} \left[S_1 + (-m + a \omega) S_0\right]\:,\\ 
    L^\dagger_2 &\left[ \rho^3 S \left( \bar{\rho}^2 \rho^{-4} \right)' \right]\Big|_{\theta=\pi/2} = \frac{2}{r^2} \left[S_1 + (-m + a \omega) S_0\right]\:,
\end{align}
\end{subequations}
where $S_0=S(\pi/2)$, $S_1=\partial_\theta S|_{\theta=\pi/2}$ are the spin-weighted spheroidal harmonics and their derivatives evaluated at $\theta=\pi/2$, while $\phi=\phi(r)$ and $t=t(r)$ are evaluated along the particle's trajectory.
In the case of an equatorially plunging particle at rest at infinity (i.e. $\Theta=0$, $\theta=\pi/2$ and $\EN=1$), the expressions above reduce to those in Ref.~\cite{Nakamura:1987zz}.

For the numerical computation of the excitation coefficients in
Eq.~\eqref{eq:excitation_coefficients}, and more generally of the waveform in Eq.~\eqref{eq:wf_SN_complete}, we need to pay attention to the asymptotic behavior of $W(r)$, appearing in the integral $I_q$, at the horizon and at the ISCO.
The function $W(r)$ is regular at the horizon, as can be seen looking at the expressions~\eqref{eq:W_}. In particular, the function $\zeta(r)$ appearing in those expressions is not divergent at the horizon, since it can be written in terms of the coordinates $v$, $\tilde\phi$ as $\zeta(r)=\omega v(r)-m\tilde\phi(r)$.
 
Let us now consider the behavior of $W(r)$ at the ISCO, $r_{\rm I}$, where the function $\mathcal{R}$ has triple roots. This leads to the divergence of some of the functions appearing in Eq.~\eqref{eq:W_}---for instance, of $h_1$ in Eq.~\eqref{eq:h_}---as $(r-r_{\rm I})^{3/2}$. However the limit $r\to r_{\rm I}$ corresponds to $t\to-\infty$ for critical plunge geodesics, and in this limit the factor $e^{i\zeta(r)t(r)}\sim e^{i\omega t(r)}$   vanishes exponentially for QNMs, which have $\Im[\omega]<0$.

When the waveform is computed for frequencies different from those of the QNMs, as in the computation of the numerical solution of the SN equation in Eq.~\eqref{eq:wf_SN_complete},  
the polynomial divergence at the ISCO leads to a divergence of the waveform. This is due to the fact that the orbiting source emits radiation starting at times $t\to-\infty$. Such a divergence does not affect the final waveform in the time domain; however, it can lead to numerical problems. For this reason, in the comparison discussed in Eq.~\eqref{eq:wf_SN_complete} we have started the integration slightly inside the ISCO, at a radius $r_p=r_{\rm I}(1-\epsilon_{\rm I})$. This approximation corresponds to starting the plunge at finite times. We chose $\epsilon_{\rm I}=0.003$, and we verified that the time-domain waveform does not change significantly when $\epsilon_{\rm I}$ is increased to $\epsilon_{\rm I}=0.01$.

\section{Supplemental results}\label{app:additional_plots}
\begin{figure}[t]    \centering\includegraphics[width=0.98\linewidth]{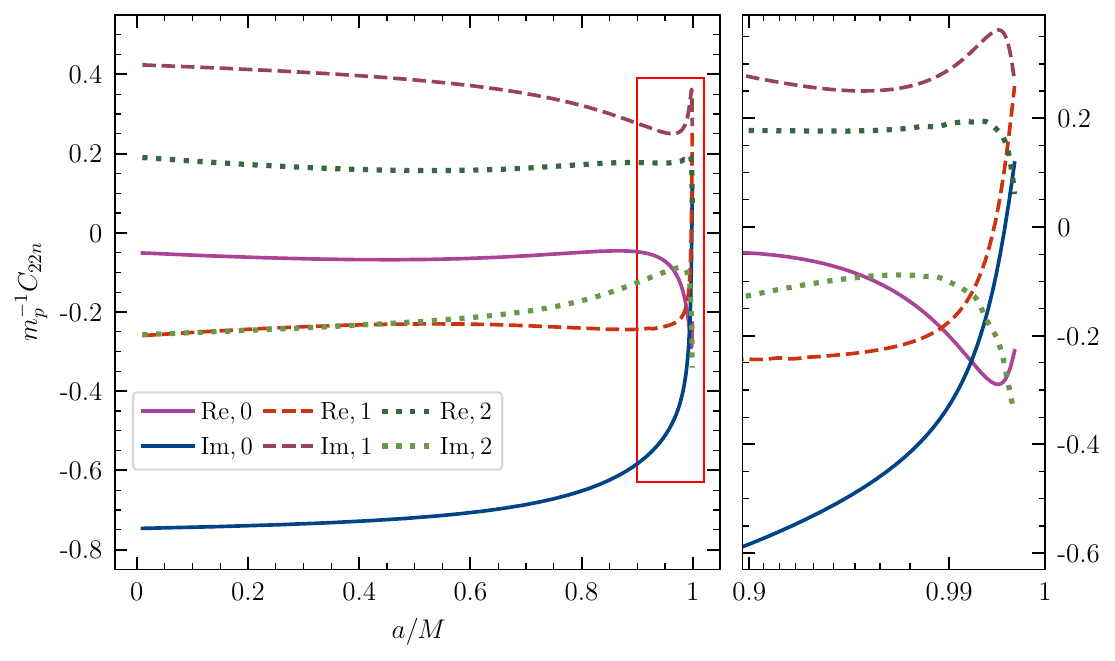}
    \caption{Real and imaginary parts of the excitation coefficients for $\ell =m=2$ and $n=0,\,1,\,2$ as functions of $a/M$. Numbers in the legend refer to the overtone number $n$.}
    \label{fig:app:exc_coeff_reim}
\end{figure}

\begin{figure}[t]
    \centering
    \includegraphics[width=0.98\linewidth]{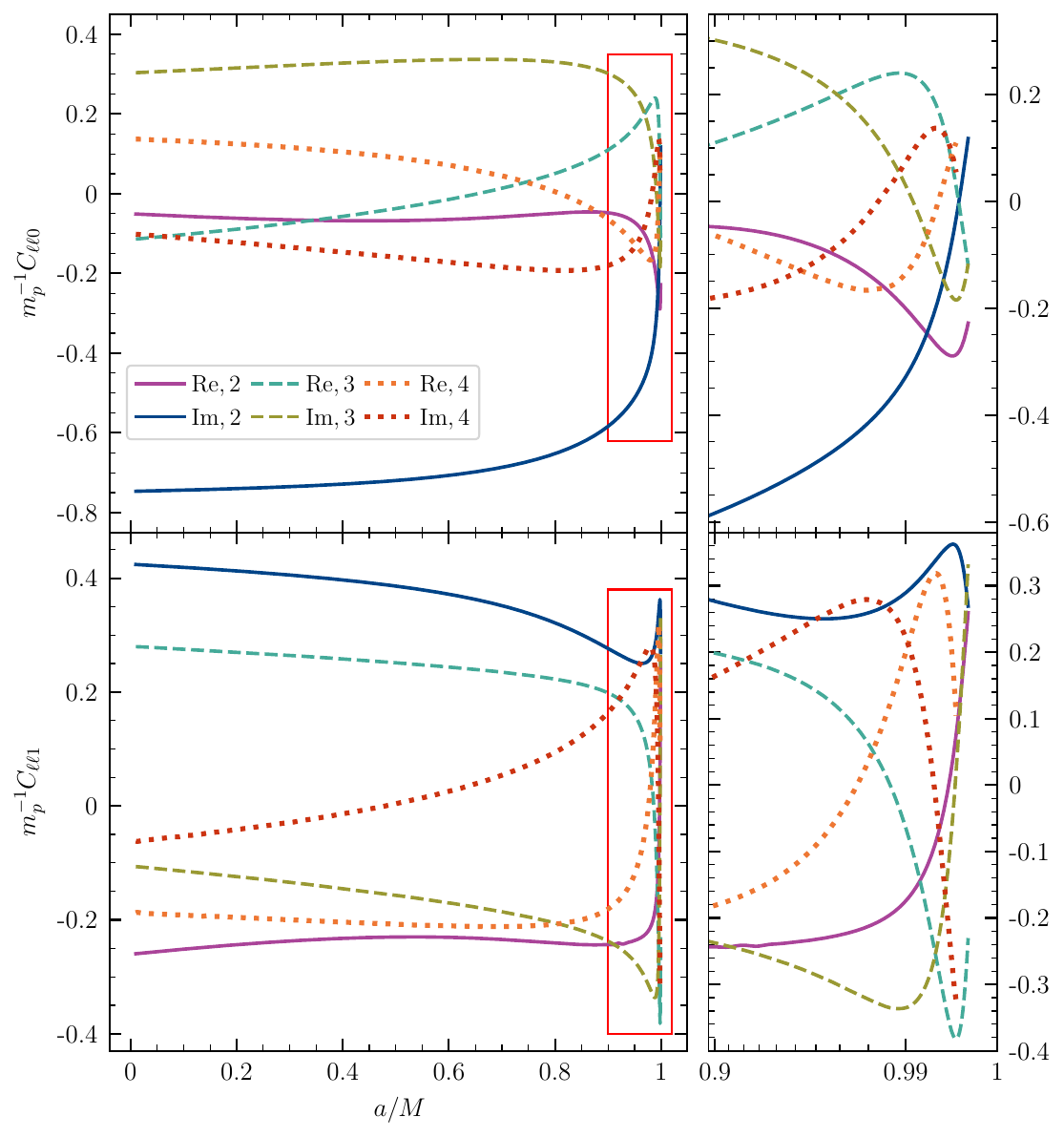}
   \caption{
   Same as Fig.~\ref{fig:app:exc_fac_HM_ReIm}, but for the excitation coefficients.}
    \label{fig:app:exc_coeff_HM_reim}
\end{figure}

For completeness, in Figs.~\ref{fig:app:exc_coeff_reim} and \ref{fig:app:exc_coeff_HM_reim} we show the real and imaginary parts of the excitation coefficients as functions of $a/M$. As discussed in Appendix~\ref{sec:excitation_factors}, they do not exhibit rapid oscillations at high spin.

\begin{table}[t]
    \centering
    \caption{Some values of $B_q$ and $C_q$. For any  $q=(\ell,m,n)$, the first (second) row refers to $a/M=0.4$ ($a/M=0.9$).}
    \begin{tabular}{c|c|c}
       $q$&$B_q$ & $C_q$  \\
    \hline
    \multirow{2}{*}{$(2,2,0)$} &$0.0932286 + 0.0690548 i$& $ -0.0677534 - 0.72847 i$ \\
   &$-0.0987088 - 0.0216979 i$ &$ -0.0478267 - 0.583508 i$\\
   \hline
    \multirow{2}{*}{$(2,2,1)$} &$0.186393 - 0.208695 i$&$-0.232469 + 0.396845 i$\\
    &$-0.0704434 + 0.574761 i $ &$ -0.243727 + 0.276398 i$\\
    \hline
    \multirow{2}{*}{$(2,2,2)$} &$-0.266837 - 0.186649 i$&$0.159947 - 0.233151 i$\\
    &$1.54187 - 0.248651 i$ & $0.177498 - 0.125972 i$\\
    \hline
    \multirow{2}{*}{$(3,3,0)$}& $-0.0291409 - 0.0941458 i$ & $-0.0568635 + 0.327784 i$\\
    &$-0.0440632 + 0.0475128 i$&$0.109696 + 0.301796 i$\\
    \hline
    \multirow{2}{*}{$(3,3,1)$}& $-0.390134 + 0.100988 i$ & $0.258141 - 0.144968 i$\\
    &$0.380624 + 0.40664 i$& $0.198459 - 0.23752 i$\\
    \hline
    \multirow{2}{*}{$(4,4,0)$}& $-0.0237187 + 0.0842478 i$ & $ 0.105021 - 0.146396 i$\\
    &$0.0205005 + 0.0517338 i$&$-0.0638536 - 0.180055 0$\\
    \hline
    \multirow{2}{*}{$(4,4,1)$}& $0.4628 + 0.145151 i$ &$ -0.204012 - 0.0140153 i$\\
    &$0.627475 - 0.194611 i$& $-0.181052 + 0.164074 i$\\
    \hline
    \end{tabular}
    \label{tab:app:results}
\end{table}
In Table~\ref{tab:app:results}, for reference, we list the values of $B_q$ and $C_q$ for selected modes and selected values of the spin.

\begin{figure}[t]
    \centering
    \includegraphics[width=0.96\linewidth]{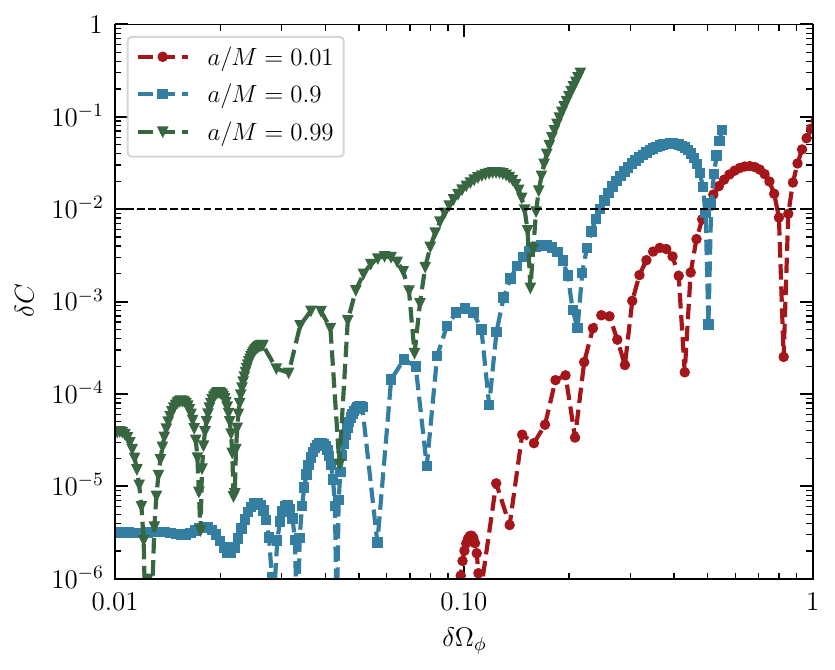}
    \caption{Relative variation $\delta C$ of the excitation coefficients of the fundamental mode as functions of $\delta\Omega_\phi$, for $a/M= 0.01,0.9,0.99$.
    The horizontal line corresponds to a $1\%$ variation.
    }
    \label{fig:orb_exc_freq}
\end{figure}

\section{Dependence of the excitation coefficients on the starting point of the plunge}\label{app:startingpoint}

In the main text we have assumed that the particle plunges following a critical trajectory that starts exactly at the ISCO. In general, standard schemes for the transition from inspiral to plunge (e.g.,~\cite{Ori:2000zn}) assume that the plunge occurs with energy $\EN= \EN_{\rm ISCO}+\delta\EN$ and angular momentum $\ANG=\ANG_{\rm ISCO}+\delta\ANG$ such that the particle is within the ISCO, i.e., $r_p=r_{\rm I}+\delta r$ with $\delta r<0 $.
In the limit $m_p/M\to 0$, the plunge reduces to a critical geodesic as $\delta\ANG=\delta \EN=\delta r\to0$. 
To consistently take into account the change in energy and angular momentum, we should consider different trajectories to describe the plunge. This is beyond the scope of this paper, but in this appendix we estimate the dependence of the excitation coefficients on the starting point of the plunge.
We focus on the excitation coefficient $C_{220}$, and study how its estimate $\hat{C}_{220}$ varies as a function of $r_p$, the upper limit of the regularized integral $\mathcal{I}_q$ in Eq.~\eqref{eq:Iq_regularized_compact}. 
In Fig.~\ref{fig:orb_exc_freq} we show the absolute value of the relative 
difference 
\begin{equation}
    \delta C=\Bigg|1-\frac{\hat{C}_{220}}{C_{220}} \Bigg|\ 
\end{equation}
as a function of the relative shift of the angular frequency of a circular orbit at radius $r_p$,
\begin{equation}
\Omega_{\phi}(r_p)=\frac{\sqrt{M}}{r_p^{3/2}+a\sqrt{M}}\,,
\end{equation}
relative to $\Omega^{\rm ISCO}=\Omega(r_{\rm I})$, i.e.,
\begin{equation}
    \delta\Omega_\phi(r_p)=\Bigg|1-\frac{\Omega_{\phi}(r_p)}{\Omega^{\rm ISCO}_{\phi}}\Bigg|\,.
    \label{eq:frequency_ratio}
\end{equation}
We plot this quantity in the range $r_p\in[r_{\rm LR},r_{\rm I})$.
From Fig.~\ref{fig:orb_exc_freq} we observe that the excitation 
coefficients are only mildly dependent on the starting point of the integration: for $a/M=0.01$, for example, $\delta C<1\%$  as long as $\delta\Omega_{\phi}\lesssim 0.1$. 
For higher spins, $\delta C<1\%$ up to $\delta\Omega_\phi\sim \mathcal{O}(1)$.
%

\bibliography{biblio}
\end{document}